\documentclass[final,1p,times]{elsarticle}

\usepackage{graphicx}
\usepackage{amssymb}

\usepackage{lineno}
\usepackage{todonotes}
\usepackage{hyperref}
\usepackage{eurosym}
\usepackage{float}
\usepackage{amsmath}
\usepackage{threeparttable}
\usepackage{pdflscape}
\usepackage{longtable}
\usepackage{colortbl}
\usepackage{tabularx}
\usepackage[numbered]{bookmark}
\usepackage{capt-of}
\usepackage{multirow}

\newcommand{\tief}[1]{\textsubscript{#1}}

\newcommand{\CO}{CO\textsubscript{2}}
\newcommand{\tCO}[1]{$#1\;t\textsubscript{CO2}$}
\newcommand{\COabat}{CO\textsubscript{2} abatement}
\newcommand{\COabcost}{CO\textsubscript{2} abatement cost}
\newcommand{\eurokWh}[1]{$#1\,\textit{\euro}\slash \mathit{kWh}$}
\newcommand{\ctkWh}[1]{$#1\,\textit{ct}\slash \mathit{kWh}$}
\newcommand{\kEUR}[1]{$#1\,\textit{k\euro}$}

\newcommand{\eurokWel}[1]{$#1\,\textit{\euro}\slash \mathit{kW}_{el}$}
\newcommand{\eurokWth}[1]{$#1\,\textit{\euro}\slash \mathit{kW}_{th}$}
\newcommand{\eurokWp}[1]{$#1\,\textit{\euro}\slash \mathit{kWp}$}
\newcommand{\eurotCO}[1]{$#1\textit{\euro}\slash \mathit{t_{CO2}}$}

\newcommand{\DeltaNPV}{$\Delta NPV$}

\newcommand{\kWp}[1]{$#1\;\mathit{kW_p}$}
\newcommand{\kWel}[1]{$#1\;\mathit{kW}_{el}$}

\newcommand{\kWth}[1]{$#1\;\mathit{kW}_{th}$}
\newcommand{\kWh}[1]{$#1\,\mathit{kWh}$}
\newcommand{\MWh}[1]{$\mathit{MWh}$}
\newcommand{\MWha}[1]{$#1\,\mathit{MWh}\slash a$}
\newcommand{\SCR}[1]{$\mathit{SCR}_{el}$}
\newcommand{\DA}[1]{$\mathit{DA}_{el}$}
\newcommand{\DSS}[1]{$\mathit{DSS}_{el}$}
\newcommand{\cacsubs}[1]{$\mathit{cac}_{subs}$}
\newcommand{\cacll}[1]{$\mathit{cac}_{ll}$}
\newcommand{\DeltaCOexport}[1]{$\Delta CO_{\mathit{2,export}}$}

\usepackage[resetlabels]{multibib} 
\newcites{SI}{SI References}    

\journal{}
\makeatletter
\def\ps@pprintTitle{%
 \let\@oddhead\@empty
 \let\@evenhead\@empty
 \def\@oddfoot{}%
 \let\@evenfoot\@oddfoot}
\makeatother
\begin{document}

\begin{frontmatter}


\title{Optimal system design for energy communities in multi-family buildings: the case of the German Tenant Electricity Law}



\author[author1]{Fritz Braeuer}
\author[author1]{Max Kleinebrahm}
\author[author1]{Elias Naber} 
\author[DTU]{Fabian Scheller}
\author[ABE]{Russell McKenna}

\address[author1]{Institute for Industrial Production (IIP), Karlsruhe Institute of Technology (KIT), Karlsruhe, Germnay}
\address[DTU]{Energy Economics and System Analysis, Division of Sustainability, Technical University of Denmark (DTU), Kgs. Lyngby, Denmark}
\address[ABE]{Chair of Energy Transition, School of Engineering, University of Aberdeen, Aberdeen, United Kingdom}

\begin{abstract}
Involving residential actors in the energy transition is crucial for its success. Local energy generation, consumption and trading are identified as desirable forms of involvement, especially in energy communities. The potentials for energy communities in the residential building stock are high but are largely untapped in multi-family buildings. In many countries, rapidly evolving legal frameworks aim at overcoming related barriers, e.g. ownership structures, principal-agent problems and system complexity. But academic literature is scarce regarding the techno-economic and environmental implications of such complex frameworks. This paper develops a mixed-integer linear program (MILP) optimisation model for assessing the implementation of multi-energy systems in an energy community in multi-family buildings with a special distinction between investor and user; the model is applied to the German Tenant Electricity Law. Based on hourly demands from appliances, heating and electric vehicles, the optimal energy system layout and dispatch are determined. The results contain a rich set of performance indicators that demonstrate how the legal framework affects the technologies' interdependencies and economic viability of multi-energy system energy communities. Certain economic technology combinations may fail to support national emissions mitigation goals and lead to lock-ins in Europe's largest residential building stock. The subsidies do not lead to the utilisation of a battery storage. Despite this, self-sufficiency ratios of more than 90\% are observable for systems with combined heat and power plants and heat pumps. Public $CO_{2}$ mitigation costs range between 147.5--272.8 \eurotCO{}. Finally, the results show the strong influence of the heat demand on the system layout.

\end{abstract}


\end{frontmatter}

\section*{Highlights}
\begin{itemize}
\item System optimization of distinct operators and consumers in multi-family buildings
\item Inclusion of Tenant Electricity Law (TEL) leads to complex investment and dispatch decisions
\item Results show highest profitability for a multi-energy combination of PV, CHP and HP
\item CHP is favoured but profits strongly depend on the heating demand of the MFB
\item TEL incentivizes energy communities but may offset national \CO{} mitigation

\end{itemize}

\begin{table}[h]							
\section*{Nomenclature}							
\centering							
							
\small							
\begin{tabularx}{\linewidth}{ l X | l X }							
\hline							
\multicolumn{2}{p{5cm}|}{\textbf{\textit{Parameters and Symbols}}}			 & 	\textbf{\textit{Index}}	 & 		\\
$\mathit{clt}$	 & 	Calendar life time	 & 	$\mathit{inv}$	 & 	Investment	\\
$\mathit{i}$	 & 	Discount rate	 & 	$\mathit{inv,fix}$	 & 	Fixed investment	\\
$\mathit{clt_{rem}}$	 & 	Remaining calendar life time	 & 	$\mathit{inv,var}$	 & 	Variable investment	\\
$\mathit{c}$	 & 	Cost	 & 	$\mathit{rem}$	 & 	Residual value	\\
$\mathit{VAT}$	 & 	Value added taxes	 & 	$\mathit{O\&M}$	 & 	Operation \& maintenance	\\
$\mathit{EF}$	 & 	\CO{} emission factor	 & 	$\mathit{grid}$	 & 	Electricity from the grid	\\
$\mathit{A_{roof}}$	 & 	Area of roof	 & 	$\mathit{ll}$	 & 	Landlord	\\
$\mathit{\Delta C_{el}}$	 & 	Energy cost savings	 & 	$\mathit{te}$	 & 	Tenant	\\
$\mathit{SCR}$	 & 	Self-consumption rate	 & 	$\mathit{th}$	 & 	Thermal	\\
$\mathit{DSS}$	 & 	Degree of self-sufficiency	 & 	$\mathit{el}$	 & 	Electric	\\
$\mathit{DA}$	 & 	Degree of electrical autonomy	 & 	$\mathit{fees}$	 & 	Fees	\\
$\mathit{BigM}$	 & 	Big number	 & 	$\mathit{pv}$	 & 	Photovoltaic	\\
$\mathit{r_{chp,min}}$	 & 	Minimum load factor of CHP	 & 	$\mathit{self}$	 & 	Self-consumed not by tenant	\\
$\mathit{h_{chp,fullload}}$	 & 	Subsidized CHP full load hours	 & 	$\mathit{chp}$	 & 	Combined heat and power	\\
$\mathit{cap_{REL,lim}}$	 & 	Capacity limit for levy exception	 & 	$\mathit{feedin}$	 & 	Feed-in tariff	\\
$\mathit{E_{REL,lim}}$	 & 	Energy limit for levy exception	 & 	$\mathit{boiler}$	 & 	Gas boiler	\\
$\mathit{COP}$	 & 	Coefficient of performance	 & 	$\mathit{gas}$	 & 	Natural gas	\\
$\mathit{D}$	 & 	Demand	 & 	$\mathit{levy}$	 & 	REL levy	\\
$\mathit{cac}$	 & 	\COabcost{}	 & 	$\mathit{M\&I}$	 & 	Metering \& invoicing	\\
$\mathit{CF}$	 & 	Cash flow	 & 	$\mathit{wo}$	 & 	Without subsidies	\\
$\mathit{r_{el}}$	 & 	Yearly electricity price change rate	 & 	$\mathit{SCP}$	 & 	Self-consumption premium	\\
$\mathit{r_{EF}}$	 & 	Yearly emission factor change rate	 & 	$\mathit{tot}$	 & 	Total amount per year	\\
$\mathit{R}$	 & 	Revenue	 & 	$\mathit{ref}$	 & 	Reference case	\\
\textbf{\textit{Sets}}	 & 		 & 	$\mathit{opt}$	 & 	Optimized case	\\
$\mathit{l}$	 & 	technology	 & 	$\mathit{EF}$	 & 	\CO{} emission factor	\\
$\mathit{a}$	 & 	year	 & 	$\mathit{subs}$	 & 	Subsidies	\\
$\mathit{t}$	 & 	hours	 & 	$\mathit{export}$	 & 	Export into grid	\\
$\mathit{rs}$	 & 	remuneration scheme	 & 		&		\\
							
\hline							
\end{tabularx}							
\end{table}							
							
\begin{table}[h]							
\section*{Nomenclature continued}							
\centering							
							
\small							
\begin{tabularx}{\linewidth}{ l X | l X }							
\hline							
\textbf{\textit{Variables}}	 & 		 & 	\textbf{\textit{Acronyms}}	 & 		\\
$\mathit{acf}$	 & 	Annual cash flow	 & 	HH	 & 	Household	\\
$\mathit{C_{inv}}$	 & 	Discounted investment/cost	 & 	PV	 & 	Photovoltaic	\\
$\mathit{NPV}$	 & 	Net present value	 & 	CHP	 & 	Combined heat and power	\\
$\mathit{cap}$	 & 	Capacity	 & 	HP	 & 	Heat pump	\\
$\mathit{E}$	 & 	Energy	 & 	HS	 & 	Heat storage	\\
$\mathit{P}$	 & 	Power	 & 	boiler	 & 	Gas boiler	\\
$\mathit{D}$	 & 	Demand	 & 	cap	 & 	Capacity	\\
$\mathit{CO_2}$	 & 	Mass/amount of \CO{} emissions	 & 	el	 & 	Electric	\\
$\mathit{\Delta CO_2}$	 & 	Abated amount of  \CO{} emissions	 & 	th	 & 	Thermal	\\
$\mathit{Q}$	 & 	Heat	 & 	RES	 & 	Renewable energy sources	\\
$\mathit{bin}$	 & 	Binary decision variable	 & 	POV	 & 	Point of view	\\
							
\hline							
\end{tabularx}							
\end{table}							
							

\section{Introduction}

On-site renewable energy generation and utilisation in buildings has been a popular topic for decades \cite{EU.2018} and remains an important element for future sustainable urban energy systems. Especially for photovoltaic (PV) installations in urban areas, large amounts of potential remain untapped. This potential is an essential part of the bottom-up approach to energy system transition, which the \citet{EU.2018b} and the \citet{EU.2019} introduced in the form of energy communities. On the European level, policymakers drafted various laws to incentivise building owners to install on-site energy generators such as PV or Combined Heat and Power (CHP) \cite{Ines.2020}. The legislation allows for a diverse variety of business models \cite{Reis.2021} with the overall goal of lowering electricity prices and expanding the share of renewable energy sources (RES). Additionally, the policies incentivise both self-generation and self-consumption to make the energy system transition more affordable for households while reducing stress on electricity grids \cite{Abada.2020}. 

On the national level, the building stock is responsible for a significant amount of greenhouse gas emissions\footnote{In Germany, the building sector makes up more than $25\%$ of greenhouse gas emissions \cite{BMU.2020}.} and thus plays a key role in reducing the building stock's energy consumption\footnote{In Germany, its number of apartments makes up $53\%$ of the building stock \cite{Bigalke.2016}.}. While the installation of renewable energy systems in single-family buildings (SFBs) is an established practice, there is still potential in multi-family buildings (MFBs). This relatively high reduction potential remains barely utilized because of diverse ownership structures and the principal-agent dilemma. Therefore, energy communities need to consider concepts beyond the conventional self-consumption-based approach. Internal revenue streams need to provide value for both the principal and the agent, i.e. the landlord and the tenant respectively. Reviewing existing policies in different European countries, multiple cash flows, energy flows, and data flows connecting various market participants lead to rather complex legal frameworks \cite{Hall.2020}. Examples of such policies are the Private Wire Network policy in the UK, the Collective Auto Consumption policy in Spain, the Post Code Rose policy in the Netherlands and the Mieterstrom policy in Germany. Finally, the multi-energy system nature of energy communities offers higher efficiencies, higher reliability, and the integration of a larger share of RES \cite{Mancarella.2014}. Nevertheless, identifying the optimal size and dispatch for a system that considers multiple energy forms and their interactions requires large amounts of computational resources. Additionally, the temporal resolution influences the precision of the results, but a high resolution is computationally expensive \cite{Heendeniya.2020}.

As outlined in the literature review in Section 2, energy communities at the building and district scale are active research areas. Despite many contributions towards optimising multi-energy systems at these scales in recent years, most studies investigate energy communities' self-consumption in a traditional sense. They do not consider the different economic situations of investor and user, and studies about novel business models are scarce \cite{Reis.2021}. This paper, therefore, aims to fill this research gap with a fundamental techno-economic analysis of optimal energy system configurations for multiple MFBs in the context of energy communities by taking the legal framework into account. 

The key contribution and novelty of this paper is the optimization of the design and operation of an energy community considering:
\begin{itemize}
    \item multi-energy forms, electricity, heat and electric mobility, and
    \item internal revenue streams, where investor and user are non-identical, and
    \item the complexity of energy communities' legal framework, and
    \item a high temporal resolution of one year in hourly time steps, and
    \item degression effects for the remuneration of self-generated electricity.
\end{itemize}

We apply the model to four MFBs in Germany and implement the German Tenant Electricity Law (TEL)\footnote{This study considers the law in force up to the 1st. of January 2021.}. Similar to other European energy community policies, the TEL supports operators (landlords) to install PV on and CHP in MFBs and profit from the on-site consumption of the electricity by their tenants instead of relying solely on feed-in tariffs. Additionally, tenants can profit from reduced energy expenditures and higher shares of renewable electricity supply. 
The optimization model, solved on a high-performance computing system, allows us to identify non-intuitive coherence among different technologies. Additionally, the model results present challenges for the optimal system design process that arise from complex internal energy and cash flows in combination with technical constraints. Subsequently, we derive policy and business implications on avoiding pitfalls in investment and operational decisions. 


The analysis is threefold: we study the combination of different technology components (component-wise analysis), the influence of different building types (building-wise analysis) and the sensitivity to policy changes (comparison of the amendment from TEL 2020 to 2021). The key performance indicators (KPI) are the Net Present Value (NPV), the technical capacities to describe the design of the energy system, measures to describe the self-consumption behaviour and grid interaction, as well as the \CO{} emissions and abatement. 

This paper is structured as follows. In Section \ref{sec_lit_review}, we summarise the existing international academic literature on energy communities and different legal frameworks concerning TEL. As a conclusion of the literature review, we identify the scientific gap addressed by the paper. Section \ref{sec_method} gives a detailed description of the current TEL, the economic conditions, and remuneration and subsidies schemes. Furthermore, Section \ref{sec.method_OptModel} lists the most relevant model equations, and Section \ref{subsec_KPI} presents this study's KPIs. Section \ref{sec_studydesign} elaborates on the design of the three main analyses, which are evaluated in Section \ref{sec_results}. The results are further discussed in Section \ref{sec_discussion} together with policy implications, methodological shortcomings and an outlook for future studies. Section \ref{sec_conclusion} ends the paper with concluding remarks.

\section{Literature review}
\label{sec_lit_review}
Different publications deal with the benefits of decentralized multi-energy systems in multi-apartment buildings. \citet{Lindberg.2016} used a mixed-integer model to investigate solutions for zero energy MFBs with 10 apartments in Germany. While a combination of CHP-system with a PV-system has been identified as a robust cost-optimal investment, the system causes large impacts on the grid in peak hours. A techno-economic analysis of energy-related active retrofitting of MFBs is presented by \citet{Fina.2019} without considering any legal constraints. A further assessment of the tenant model in an Austrian MFB showed that the economic viability of PV-systems strongly depends on the retail electricity prices \citep{Fina.2018}. \citet{Fina.2018} conclude that the case should be a win-win situation for both landlords and tenants in Germany due to the higher retail prices, but the Austrian TEL is also less restrictive. This also applies to the model of \citep{Ferrara.2018}, which optimized the energy system design of a six-floor MFB in Northern Italy. A similar study of a multi-apartment building with a stronger focus on energy autonomy and hybrid systems has been conducted by \citet{COMODI2015854}. While the authors conclude that a battery would increase independence from the grid, the costs are very  high. \citet{PALOMBA2020190} optimised the supply systems of two main MFB typologies in three climate regions in Europe. The results showed the possibility of achieving high solar fractions for domestic hot water even in northern climates. 
By taking the main German levies and taxes for self-generation and -consumption into account, the optimization results of \citet{McKenna.2017} indicate a shift in the economically optimal level of electrical self-sufficiency with scale. The effect of the then current legal regime on the optimal design of a CHP is also highlighted by \citet{merkel2017modellgestutzte}. 

Research with a sharper focus on the national TEL is mostly presented in non peer-reviewed (grey) literature and/or written in  German. Three simple exemplary TEL business-case evaluations in terms of a 20 party MFB are presented in the German study of \citet{Harder.2020}. While the installation of a CHP-system leads to an economic advantage of around 600 \euro{} per annum, the PV-system on the other hand only leads to around 40 \euro{} per annum. A combination of the two systems with simultaneous consideration of e-mobility loads demonstrates an economic advantage of around 50 \euro{} per annum. The implementation of the additional battery system leads to higher security but also a lower economic advantage \citep{Harder.2020}. Additionally, the non peer-reviewed article by \citep{Manns.2019} applies a techno-environmental analysis of MFBs with PV-systems and electric vehicle charging stations under the TEL. It is shown that 2-4 tCO2e/a emissions can be saved, depending on the building size. The economic advantages of the TEL for the tenant and the operator is also supported by the results of \citet{Scheller.2017} and the conference proceeding of \citet{Seim.2017}. Furthermore, the practical tool of \citet{Knoop.2018} represent a simple assessment tool for the TEL in Germany. An overview of existing policies in the EU that foster the potential of collective RE prosumers in energy communities is given by \citet{Ines.2020}. Detailed examples of practical implementations of different forms of the TEL in the UK (Private Wire Networks policy), Spain (Collective Auto Consumption policy), Netherlands (Post Code Rose policy) and Germany (Mieterstrom policy) are presented by \citet{Hall.2020}. 

While the number of publications indicates an increasing interest in energy communities or neighbourhoods, the legal frameworks in action are considered to be at an early stage \citep{Frieden.2019}. 
The energy hub approach has been taken into account by various publications to determine the optimal energy system design of neighbourhoods. An overview is given in the review article by \citet{Mohammadi.2017}, who lists 129 scientific papers using the approach; most of them without considering the legal framework apart from the feed-in tariffs. Thereby, \citet{Ghorab.2019} implement a multi-objective optimization model to minimize the overall cost and emissions of different building archetypes in Canada forming an energy hub. A similar approach has been used by \citep{Orehounig.2015} to evaluate and size neighbourhood energy systems according to their energy-autonomy, economic and ecological performance. A multi-objective optimization model for the investment planning of distributed heat and electricity supply systems has been applied by \citep{Falke.2016} to a district with various apartment buildings in Germany. 
\citet{Batic.2016} present a linear programming approach to optimize the daily schedule of a multi-energy system on a research campus in Belgrade. The focus is on demand-side management applications considering a dynamic electricity tariff. \citet{Ma.2018} simulate the operation of an energy district with 1000 buildings in China to increase self-consumption. \citet{Jing.2020} present a general market concept for local energy hubs in China connecting residential and commercial prosumers. For the optimization of a multi-energy system in a building in Tehran, \citet{Eshraghi.2019} formulate an optimization model considering a flexible electricity tariff. Scheller et al. \citep{Scheller.2020,Scheller.2018b,scheller2018towards} assess community business models in Germany with and without legal aspects. They propose an optimization model to determine the optimal investment and operation of a community electricity storage system. 

The foregoing discussion highlights previous attempts to optimise multi-energy system designs. While individual publications take different legal conditions of decentralised technologies into account, none of the techno-economic studies properly consider energy communities, where the operator and the consumer are not the same entity as is the case in MFBs. In such a case, there is a variety of legal requirements regarding the market remunerations and premiums as well as the statutory fees, levies and taxes \citep{Scheller.2018,Harder.2020}. While \citet{Reis.2021} elaborated an overview of archetypes covering the wide range of conceptual possible business models in energy community settings, the analysis revealed that traditional self-consumption place-based communities are still dominating the research landscape. Business models involving differentiated services and non-identical investor and users are still scarce. Furthermore, as our analysis showed, hardly any techno-economic study has investigated the complexity of the legal setting in general and the German TEL in MFBs in particular. To fill this gap, the economic and environmental implications need to be assessed by addressing the uncertainties. The current contribution, an extension of \citet{Braeuer.2019}, develops and applies a techno-economic optimization model for MFBs with a special focus on the TEL in Germany. In addition to the legislative texts \citep{EEG.2017a,EEG.2021}, the legal opinions of \citet{Herz.2018} and the practical descriptions of \citet{Behr.2017} serve as a basis for the implementation. The results provide a more detailed view on single archetypes of emerging energy community arrangements as conceptually analyzed by \citep{Reis.2021} to identify an optimal solution for arrangements with non-identical operator and end-consumer. In this context, this study can be seen as a first attempt for a more in-depth analysis of different European energy community archetypes.

\section{Methodology}
\label{sec_method}
For this study, we apply a mixed-integer linear optimisation model (MILP) that incorporates the legal frameworks of the TEL 2021. First, the regulatory framework and the monetary flows of the legal frameworks relevant for energy system modelling are presented in Section \ref{subsec_reg_Fra}. The implementation of the regulatory frameworks in the MILP model is explained in Section \ref{sec.method_OptModel}. Finally, the key performance indicators required for evaluating the model results are shown in Section \ref{subsec_KPI}.

\subsection{Implementation of the regulatory framework}
\label{subsec_reg_Fra}
The Tenant Electricity Model (TEM) is an established concept that enables self-consumption of self-generated electricity in an MFB. The Renewable Energies Law (REL) in 2014 \citep{EEG.2017a} together with the Combined Heat and Power Law (CHPL) \cite{KWKG.2018} initialised the Tenant Electricity Law (TEL), and the most recent changes in the REL 2021 \citep{EEG.2021} and CHPL 2020 \cite{KWKG.2020} introduce substantial modifications to the levels of the subsidies. The TEL aims to incentivise the installation of RE systems and self-consumption in MFBs. The law establishes the legal framework for the landlord, leaseholder, or contractor to act as the tenant's electrical contractor. The electricity tariff must be at least $10\%$ less than the local basic electricity provider's tariff. The contract must be renewed every year, and it must not have any influence on the rent agreement. \citet{Behr.2017} provide a handbook on the initial TEL. 

\begin{figure}[h]
\includegraphics[width=\linewidth]{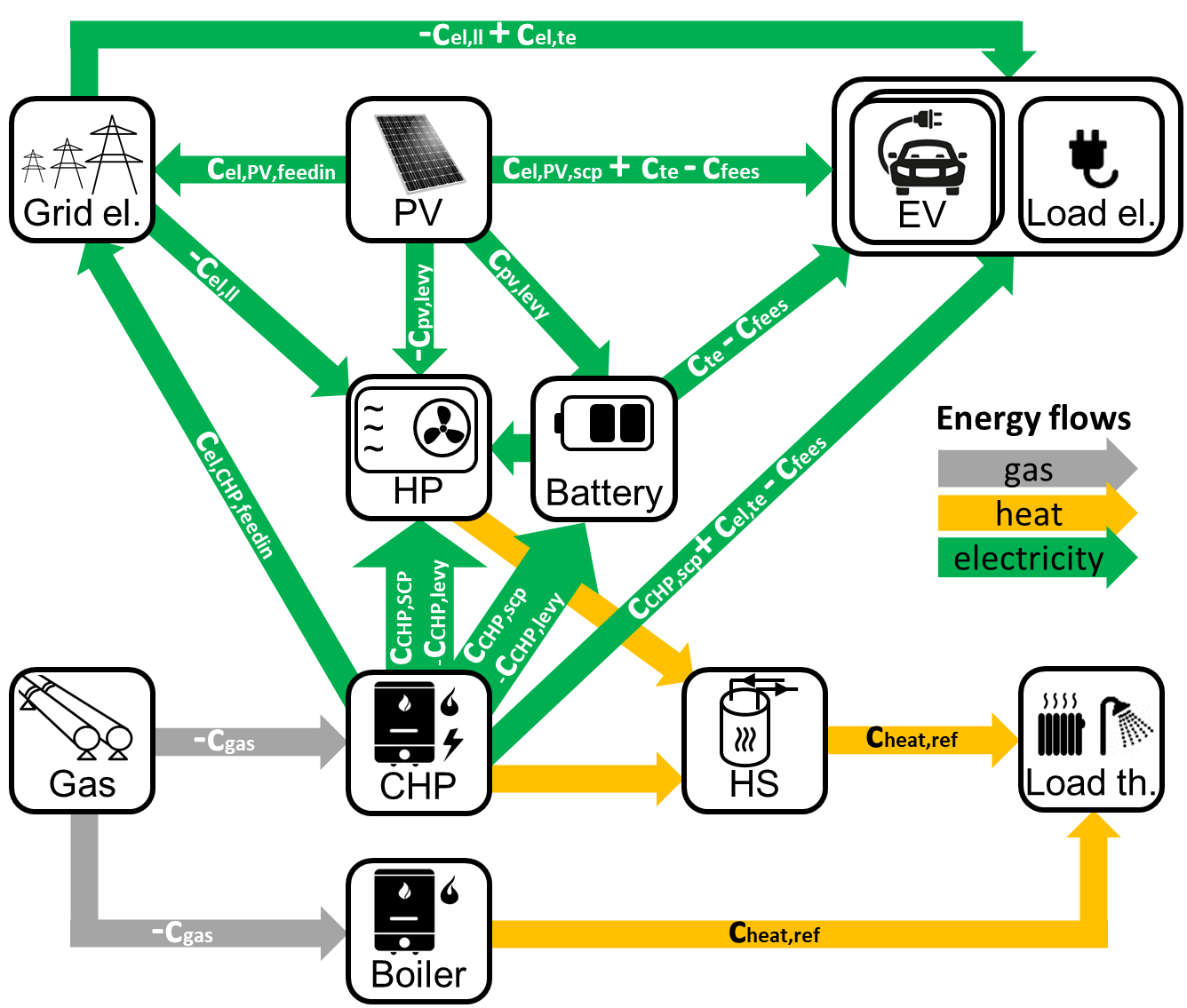}
\caption{Illustration of the Tenant Electricity Model (TEM) and the optional technological components. The possible energy flows are illustrated by arrows and the symbols represent the remuneration for the respective energy flow from the landlord's point of view.}
\label{fig.cashflow_model}
\end{figure}

The legal framework incentivises a high amount of self-generated and self-consumed electricity by guaranteeing different subsidy payments per kilowatt hour or levy and fee exceptions. To better illustrate the remunerations for different energy flows, Figure \ref{fig.cashflow_model} presents a TEM in an MFB, its optional technological components and the possible energy flows as arrows. The symbols in the arrows indicate the specific payment linked to each energy flow and are further explained in Section \ref{sec.method_OptModel}. 

\citet{Herz.2018} summarise the remuneration concept of PV-electricity in a TEM according to the REL 2017. Even though the REL 2021 changes the amount of compensation, the underlying scheme remains the same. In the REL of 2021, for all PV systems smaller than \kWp{100}, self-consumed PV electricity sold to the tenants is exempted from all fees except the REL levy, value-added taxes, and costs for metering \& invoicing. Additionally, a Tenant Electricity Premium or a so-called Self-Consumption Premium (SCP) is remunerated. Surplus PV electricity can be fed into the grid under a feed-in tariff. Self-consumed PV electricity directed to an HP or battery does not receive an SCP but is only charged with a REL levy for systems larger than \kWp{30} and \MWha{30}. Notably, the amount of feed-in tariff and SCP depends on the installed PV capacity, see Section \ref{sec_app_PVremuneration}. The PV subsidy scheme is fixed for 20 years. 

A similar concept with varying premiums and tariffs holds for self-generated electricity from a CHP unit. For an up-to-date summary of the most recent CHPL 2020 implications, see \citet{UBA.2020}. In the CHPL 2020, a de-minimis limit exempts CHP-units with a size smaller than \kWel{10} and a self-consumption of less than \MWha{10} from the REL levy. Larger units have to pay $40\%$ of the levy on self-consumed electricity. In contrast to the PV remuneration, feed-in tariff and SCP are fixed independent of size. Additionally, the SCP is paid for CHP electricity sold to tenants and self-consumed in an HP or battery. The described subsidy scheme is limited to units smaller than \kWel{50} and 30,000 full load hours.

How to consider the cost for heat generation from a heat pump in a TEM and pass it on to the tenants is an open question. A consultation of the authors with a German law firm specializing in energy law revealed no precedents for such a model. For this study, the authors consider a heat pump as a self-consuming unit. Thus, self-generated CHP electricity drawn by the heat pump is remunerated with a SCP. Nonetheless, the overall assumption is that the switch to a TEM does not inflict additional costs for heating on the tenants. It is priced in the same way as in the reference system. Here the operational cost is the gas expenditure for a gas boiler.

\begin{figure}[h]
\includegraphics[width=.6\linewidth]{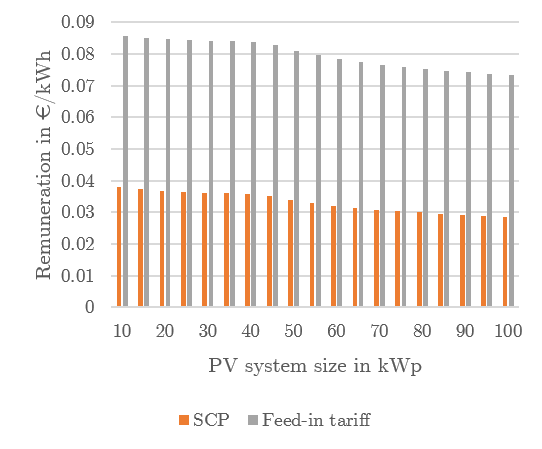}
\hspace{2pc}
\begin{minipage}[b]{10pc}
\caption{Feed-in tariff and self-consumption premium depending on remuneration scheme of discrete PV system sizes}
\label{fig.feedinpremium}
\end{minipage}
\end{figure}

\subsection{Optimization model}
\label{sec.method_OptModel}

The model used to determine the energy system of the MFB is based on \cite{Kleinebrahm.2018}. The MILP model defines the optimal system set up and operation from a landlord's point of view (POV). The model determines the optimal energy flows on an hourly basis for one representative year matching the households' electricity and heating demand. Electricity tariffs and surcharges change with a yearly rate ($r_{el}$). The objective function is the maximisation of the Net Present Value (NPV). The decision variables are binary variables for different technology options. Furthermore, positive continuous variables describe the technology dimensions, namely installed capacity and the energy flow between them. One notable exception is the CHP capacity; it is given as an exogenous model parameter further explained in Section \ref{sec_setupcompwise}. The model is implemented in Matlab and solved with the CPLEX solver. Figure \ref{fig.cashflow_model} gives an overview of the considered system components, the energy and cash flows. 

The objective function in Equation \ref{eq.obj} describes the \textit{NPV} as the the difference between the discounted investment ($C_{inv}^l$), which considers future re-investment and residual value, and the discounted annual cash flow ($\mathit{acf}$). The discounted investment is shown in Equation \ref{eq.inv} and considers the initial investment in the year $a=0$ for each technology $l$. The initial investment consists of a fixed component ($c_{inv,fix}^{l,a=0}$) for every technology option ($bin_{fix}^l$) and a variable component ($c_{inv,var}^{l,a=0}$) that depends on the installed capacity ($cap^l$). Additionally, Equation \ref{eq.inv} includes a discounted reinvestment after the calendar lifetime ($\mathit{clt}$) of technology $l$ is reached and a discounted residual value after the considered investment period $A$. This allows the different lifetimes of the technology options to be taken into account. 

Equation \ref{eq.acf} represents the annual cash flow illustrated in Figure \ref{fig.cashflow_model}, consisting of five parts. The first part considers a term for operation and maintenance, which is a percentage of the installed capacity ($cap^l$) for each technology \textit{l}. The second part defines the revenue from the self-generated electricity (\textit{P}) the landlord (\textit{ll}) sells to the tenants (\textit{te}). The third part determines the revenue and expenditure on electricity drawn from the grid. The fourth and fifth parts define the additional revenue from PV and CHP generated electricity respectively. The last part considers the revenue from providing heat to the tenants.

In more detail, the landlord sells the self-generated electricity to the tenants ($P_{ll,te}^t$) for every hour \textit{t} for the tenant electricity price ($c_{el,te}^a$) avoiding various fees ($c_{fees}^a$). The avoided fees are shown in Equation \ref{eq.fees}. Concerning grid electricity ($P_{grid,ll}^t$), the landlord pays the price $c_{grid,ll}^a$. Grid electricity that is directly forwarded to the tenants ($P_{grid,te}^t$) is charged with the tenant electricity price and not exempted from any fees.

Next to selling self-generated electricity for the tenant electricity price, additional revenue from PV electricity is either generated by feeding electricity into the grid ($P_{pv,grid}^{t,rs}$) with a fixed feed-in tariff ($c_{pv,feedin}$) or by the SCP ($c_{pv,scp}$) paid on tenant electricity from PV ($P_{pv,te}^{t,rs}$). PV electricity that is not self-consumed by the tenants but within the building ($P_{PV,self}^{t,rs}$), for example in a central heat pump or battery storage system, is charged with additional surcharges or levies ($c_{levy}^a$).

The German legislation makes the remuneration value dependent upon the installed PV capacity. To consider this dependency, we introduce remuneration schemes that result in discrete remuneration steps for PV electricity, further explained in Section \ref{eq_app_PVremuneration}. For this study, the spectrum of possible remuneration schemes (\textit{rs}) is divided into 19 sectors as illustrated in Figure \ref{fig.feedinpremium}. Every remuneration scheme coincides with a respective value of the feed-in tariff, the tenant electricity premium and the self-consumption levies. With Equations \ref{eq.bin_PV}, \ref{eq.cap_PV}, and \ref{eq.P_PV_int} the model selects the relevant remuneration scheme. The binary variable $\mathit{bin}_{pv}^{rs}$ indicates the selected revenue scheme and Equation \ref{eq.bin_PV} states that the model can only choose one \textit{rs}. Considering $\mathit{bin}_{pv}^{rs}$, Equation \ref{eq.cap_PV} defines the variable for the PV capacity ($cap_{pv}$) to be less or equal to upper level in the selected remuneration scheme ($cap_{pv}^{rs}$). The PV electricity flow ($P_{pv,cf}^{t,rs}$) is defined for the full set of remuneration schemes. Nonetheless, Equation \ref{eq.P_PV_int} allows only $P_{pv,cf}^{t,rs}$ from the selected remuneration scheme to be greater than zero\footnote{\textit{BigM} describes a significantly large number according to the Big M Method in operations research.}. Equation \ref{eq.P_PV_int} is defined for the respective cash flows (\textit{cf}) tenant electricity (\textit{te}), feed-in (\textit{feedin}) or REL levy (\textit{self}).


In analogy to PV electricity, the fifth part of the annual cash flow in Equation \ref{eq.acf} describes the additional revenue for CHP electricity. Electricity generated by the CHP unit can either be fed into the grid ($P_{chp,grid}$), remunerated by the feed-in tariff, sold as tenant electricity ($P_{chp,te}$) or self-consumed in a heat pump or battery storage system ($P_{chp,self}$). The latter two options profit from the SCP ($c_{chp,scp}$). Electricity that is self-consumed but not by the tenants ($P_{chp,self}$ and $P_{chp,self,wo}$) is charged with a levy ($c_{levy}$). Additionally, Equation \ref{eq.acf} considers the expenditure for gas consumption of the CHP unit as the quotient of the generated electricity ($P_{el,chp}^t$) and the electric efficiency ($\eta_{el,chp}$) times the gas price ($c_{gas}^a$). Equation \ref{eq.CHP_Pel} states that the CHP generated electricity can either be remunerated by the subsidy scheme or it is distributed without any additional payments (marked by the index \textit{wo}) except the tenant electricity price. Equation \ref{eq.CHP_Pmin}, Equation \ref{eq.CHP_Pmin2} and Equation \ref{eq.CHP_fullload} constrain the CHP operation. The former two define the electrical output of the CHP unit ($P_{el,chp}^t$) as a semi-continuous variable to be either zero or greater than a minimal output ($P_{chp,min}$), which is set as a factor ($r_{chp,min}$) of the installed capacity, Equation \ref{eq.CHP_Pmin2}. The latter Equation \ref{eq.CHP_fullload} restricts the the operational full load hours per year ($\frac{\sum_t^T P_{el,chp}^t}{cap_{chp}}$) to remain below a fixed limit ($h_{chp,fullload}$). 

The German legislation states that the levy on self-consumed CHP electricity is only paid once the installed capacity or the produced energy amount exceeds a trivial limit ($cap_{chp,lim}$ or $P_{chp,lim}$ respectively). Equation \ref{eq.CHP_self_REL} introduces the binary variable $bin_{chp,lim}$ that equals one once the model chooses to exceed the specified capacity limit. Equation \ref{eq.CHP_P_self_REL} defines the self-consumed electricity amount that is charged with the additional levy once either limit is exceeded.

The final part of Equation \ref{eq.acf} describes the revenue from satisfying the heat demand of the tenants. Heat is sold to the tenants for a price that equals the operational costs of the gas boiler. Heat is provided either through the CHP, the boiler, the heat pump or a combination of all three. Equation \ref{eq.Q_HP} states that the heat pump converts electricity coming either from the PV system, the CHP or from the grid with a time- and temperature-dependent COP into heat. The self-generated electricity is considered as part of the self-consumption electricity flow $P_{chp,self}^t$ and $P_{pv,self}^t$ in Equation \ref{eq.acf}. Further explanations of the model can be found in the Appendix.

\begin{equation}
\label{eq.obj}
\max \mathit{NPV}, \mathit{NPV}=-\sum_{l\in L}C_{inv}^l+\sum_{a=0}^{A}\frac{\mathit{acf}^a}{(1+i)^a}
\end{equation}

\begin{equation}
\label{eq.inv}
C_{inv}^l=c_{inv,fix}^{l,a=0}\cdot bin_{fix}^l + c_{inv,var}^{l,a=0}\cdot cap^l + \frac{c_{inv,var}^{l,a=clt^l} \cdot cap^l}{(1+i)^{clt^l}} - \frac{clt_{rem}^l}{clt^l} \cdot \frac{c_{inv,var}^{l,a=A} \cdot cap^l}{(1+i)^A}
\end{equation}

\begin{equation}
\label{eq.acf}
\begin{split}
\mathit{acf}^a= & \sum_{l=1}^L -c_{\mathit{O\&M}}^l\cdot cap^l 
+\\
& \sum_{t=1}^{8760}\bigg(P_{ll,te}^{t}\cdot (c_{el,te}^{a}-c_{fees}^{a}) +\bigg[-P_{grid,ll}^{t}\cdot c_{el,ll}^{a}+P_{grid,te}^{t}\cdot c_{el,te}^{a}\bigg]
+\\
& \bigg[\sum_{rs}^{19} \bigg(P_{pv,grid}^{t,rs}\cdot c_{pv,feedin}^{rs}+ P_{pv,te}^{t,rs}\cdot c_{pv,scp}^{rs} - P_{pv,self}^{t,rs}\cdot c_{pv,levy}^{a,rs} \bigg)\bigg]
+\\
& \bigg[P_{chp,grid}^{t}\cdot c_{chp,feedin} + (P_{chp,te}^{t} + P_{chp,self}^{t}) \cdot c_{chp,scp} 
- \\ 
& (P_{chp,self}^{t} + P_{chp,self,wo}^{t}) \cdot c_{levy}^a - P_{chp,el}^{t}\cdot \frac{c_{gas}^{a}}{\eta_{chp,el}} \bigg]
+\\
& (Q_{te}^{t} - Q_{boiler}^t) \cdot \frac{c_{gas}^{a}} {\eta_{boiler}}
\bigg) \;\; \forall \, a \in A
\end{split}
\end{equation}

\begin{equation}
\label{eq.fees}
c_{fees}=c_{levy}^a+\mathit{VAT}+c_{M\&I}^a
\end{equation}

\begin{equation}
\label{eq.bin_PV}
\sum_{rs}^{19} bin_{pv}^{rs} = 1
\end{equation}

\begin{equation}
\label{eq.cap_PV}
cap_{pv} \leq \sum_{rs}^{19} bin_{pv}^{rs} \cdot cap_{pv}^{rs}
\end{equation}


\begin{equation}
\label{eq.P_PV_int}
P_{pv,cf}^{t,rs} \leq bin_{pv}^{rs} \cdot bigM \;\;\forall\,rs, cf \in \{\mathit{feedin,scp,levy}\}
\end{equation}

\begin{equation}
\label{eq.CHP_Pel}
P_{el,chp}^t = P_{chp,grid} + P_{chp,te} + P_{chp,self} + 
                P_{chp,grid,wo} + P_{chp,te,wo} + P_{chp,self,wo} 
\end{equation}

\begin{equation}
\label{eq.CHP_Pmin}
bin_{chp}^t \cdot P_{chp,min} \leq P_{el,chp}^t \;\; \forall \,t \in T
\end{equation}

\begin{equation}
\label{eq.CHP_Pmin2}
P_{chp,min} = r_{chp,min} \cdot cap_{chp}
\end{equation}

\begin{equation}
\label{eq.CHP_fullload}
\frac{\sum_t^T (P_{chp,grid}^t + P_{chp,te}^t + P_{chp,self}^t)}{cap_{chp}} \leq h_{chp,fullload}
\end{equation}

\begin{equation}
\label{eq.CHP_self_REL}
cap_{chp} \leq cap_{lim,REL} + bin_{chp,levy} \cdot \mathit{BigM}
\end{equation}

\begin{equation}
\label{eq.CHP_P_self_REL}
\sum_t^T \bigg( P_{chp,self,hp}^t + P_{chp,self,bat}^t - P_{chp,self}^t - P_{chp,self,wo}^t \bigg) \leq E_{chp,lim,levy} \cdot (1 - bin_{chp,levy} )
\end{equation}



\begin{equation}
\label{eq.Q_HP}
Q_{hp}^t = \big[ P_{pv,self,hp}^t + P_{chp,self,hp}^t + P_{grid,hp}^t \big] \cdot \mathit{COP}^t \;\; \forall \,t \in T
\end{equation}

\subsection{Key performance indicators}
\label{subsec_KPI}
To evaluate the economic performance of the system layout, the NPV over a period of 20 years is used. To assess the self-generated energy usage the electrical self-consumption rate ($SCR_{el}$), the degree of electrical self-sufficiency ($DSS_{el}$) as well as the degree of electrical autonomy ($DA_{el}$) are employed \cite{McKenna.2017}, see Equation \ref{eq_selfcons}, \ref{eq_selfcov} and \ref{eq_autonomy}. Additionally, the Grid Interaction Index \ref{eq_gii} and the normalized Grid Interaction Index \ref{eq_gii_norm} are calculated according to \citep{McKenna.2017}.

\begin{equation}
\label{eq_selfcons}
SCR_{el}=\frac{\sum_{t}^T (P_{pv,self}^t + P_{pc,te}^t + P_{chp,self}^t + P_{chp,te}^t)}{\sum_{t}^T(P_{pv,gen}^t+P_{chp,gen}^t)}
\end{equation}

\begin{equation}
\label{eq_selfcov}
DSS_{el}=\frac{\sum_{t}^T (P_{pv,self}^t + P_{pv,te}^t + P_{chp,self}^t + P_{chp,self}^t )} {\sum_{t}^T(D_{el,te}^t + D_{el,hp}^t + D_{el,ev}^t)}
\end{equation}

\begin{equation}
\label{eq_autonomy}
DA_{el}=\frac{DSS_{el}}{SCR_{el}} = \frac{\sum_{t}^T(P_{pv,gen}^t+P_{chp,gen}^t)}{\sum_{t}^T(D_{el,te}^t + D_{el,hp}^t + D_{el,ev}^t)}
\end{equation}

\begin{equation}
\label{eq_gii}
GII=\sqrt{\frac{1}{T-1}\sum_{t}^T\left(\frac{P_{intogrid}^t}{max|P_{intogrid}^t|}-\frac{1}{T}*\sum_{t=T}^N\frac{P_{intogrid}^t}{max|P_{intogrid}^t|}\right)^2}
\end{equation}

\begin{equation}
\label{eq_gii_norm}
GII_{norm}=\frac{GII}{
\sqrt{\frac{1}{T-1}\sum_{t}^T\left(\frac{D_{el}^t}{max|D_{el}^t|}-\frac{1}{T}*\sum_{t}^T\frac{D_{el}^t}{max|D_{el}^t|}\right)^2}}
\end{equation}

For the assessment of Greenhouse Gas (GHG) emissions, only \CO{} is considered and allocated according to the 'polluter pays principle' as described in \cite{BMWi.2020}. Namely, this includes emissions from grid electricity and natural gas consumption but no emissions from PV electricity. The MFB represents the system boundaries. 

Equation \ref{eq_m_CO2_2} defines the amount of \CO-emissions ($CO_{2,i}$) for both cases $i$. It is the sum of the \CO{} emissions from electricity of the grid and \CO{} emissions from natural gas consumption over the respective investment period $T$. \CO{} emissions from the grid are the product of the total energy demand from the grid for case $i$ for one year ($D_{grid,tot}^i$) multiplied by the \CO{} emission factor of the grid ($EF_{grid}$). The increasing share of RES in the national electricity mix will further diminish $EF_{grid}$ in the future. To depict this development, $EF_{grid}$ is reduced every year by the \CO{} reduction factor ($r_{EF,grid}$). The \CO{} emissions from gas consumption are comprised of the yearly gas demand ($D_{gas,tot}^i$) for case $i$ multiplied by the \CO{} emission factor of gas ($EF_{gas}$), which is assumed not to change over the respective investment period.
  
Equation \ref{eq_CO2_abat} defines the \CO{} abatement ($\Delta CO_2$). $\Delta CO_2$ is the difference between the emitted amount of \CO{} in a reference case (\textit{ref}) and the optimized tenant electricity case (\textit{opt}).



Finally to account for emissions from exported energy two values are calculated. $CO_{2,export}$ provides the total quantity of exported emissions and $\Delta CO_{2,export}$ describes the added or reduced quantity of exported emission in comparison to the respective grid emissions. 
Equation \ref{eq_CO2_intogrid} determines $CO_{2,export}$ as the sum of the electricity fed into the grid by the respective energy generator multiplied by the respective emission factor\footnote{For further explanation of $EF_{chp,el}$ as the emission factor of a co-generation process, see \cite{AGEB.2019} and Section \ref{App_emissionsintogrid}.}, $EF_{pv}$ or $EF_{chp,el}$. 

Equation \ref{eq_CO2_intogrid2} calculates $\Delta CO_{2,export}$ as the difference between the exported emissions and the replaced emissions caused by the energy mix in the grid determined by the grid emission factor. A negative value for $\Delta CO_{2,export}$ illustrates that the feed-in electricity reduces the grid emissions and vice-versa. 
Furthermore, Equation \ref{eq_CO2_abatcost} assesses the \COabcost{} from the government's point of view ($\mathit{cac_{subs}}$) considering the paid subsidies. Equation \ref{eq_CO2_abatcost} relates the cash flow of these public subsidies ($CF_{sub}$), namely the SCP and feed-in tariff, to the abated \CO{} emissions. 


\begin{equation}
\label{eq_m_CO2_2}
\begin{split}
CO_{2,i}=\sum_a^{A}\Bigg( & D_{grid,tot}^a \cdot 
            \Big(EF_{grid}*(1-r_{EF,grid})^a\Big)
            + \\
            & D_{gas,tot}^a \cdot EF_{gas}
            \Bigg) \;\forall i \in \{ref,opt\}      
\end{split}
\end{equation}

\begin{equation}
\label{eq_CO2_abat}
\Delta CO_{2}=CO_{2,ref}-CO_{2,opt}
\end{equation}


\begin{equation}
\label{eq_CO2_intogrid}
\begin{split}
CO_{2,\mathit{export}} = \sum_t^T \Bigg( P_{pv,grid}^t \cdot EF_{pv} + \left(P_{chp,grid}^t + P_{chp,grid,wo}^t \right) \cdot 
EF_{chp,el} \Bigg) \cdot A
\end{split}
\end{equation}

\begin{equation}
\label{eq_CO2_intogrid2}
\begin{split}
\Delta CO_{2,\mathit{export}} = & \; CO_{2,\mathit{export}} - \sum_a^A \sum_t^T \Bigg( P_{pv,grid}^t \cdot 
EF_{grid} \cdot (1 - r_{EF,grid})^a \\
& + \left(P_{chp,grid}^t + P_{chp,grid,wo}^t \right) \cdot 
EF_{grid} \cdot (1 - r_{EF,grid})^a \Bigg)
\end{split}
\end{equation}

\begin{equation}
\label{eq_CO2_abatcost}
cac_{subs}=\frac{\sum_t^T CF_{subs}^t}{\Delta CO_{2}-\Delta CO_{2,\mathit{export}}}
\end{equation}


\section{Study design}
\label{sec_studydesign}

In this section, we describe the various configurations that the model is applied to. First, we briefly summarize the input data and assumptions, Section \ref{sec_inputdata}. Secondly, we differentiate between a component-wise and building-wise analysis and elaborate on the computational framework conditions, Section \ref{sec_setupcompwise}.  

\subsection{Input data and assumptions}
\label{sec_inputdata}

The baseline date for remuneration and consumer prices in our study is the 1st of January 2021. This includes the most recent changes in the REL 2021 and CHPL 2020. The assumed technology prices are based on the year 2018 and adapted by a technology-specific growth factor. The input parameters are shown in the Appendix Table \ref{tbl_inputdata1}, Table \ref{tbl_inputdata2}, and Table \ref{tbl_PV_remuneration}. Table \ref{tbl_consumerprices} shows the consumer prices for the electricity for tenant and landlord as well as the gas price. Prices are based on statistical values for Germany in 2020 \cite{Bundesnetzagentur.2020b}. They consider a growth rate of $2\%$ as well as an additional \CO-price according to \cite{BEHG.2019}, also shown in Table \ref{tbl_consumerprices}.
The feed-in tariff and self-consumption premium for PV is shown in Figure \ref{fig.feedinpremium} and Table \ref{tbl_PV_remuneration}.

The building is assumed to be in Karlsruhe, Germany, TRY-region 12 (test reference weather year of the DWD). We compare four building types. The respective heating and electricity demand profiles are derived from the Synpro tool \cite{Fischer.2015}, which considers an individual profile for each apartment within the MFB. The building details are shown in Table \ref{tbl_inputdata_bldgtypes}.
To consider electric vehicles (EV), we incorporate a variety of driving profiles derived from \cite{Kaschub.2016} based on data of the German Mobility Panel (MOP). This dataset gathers a variety of driving patterns observable for German personal motorized vehicles. With the assumption of EV characteristics for a sample of car models of different manufacturers, \cite{Kaschub.2016} translates these driving patterns into EV charging profiles. The EV-profiles are for one typical week; we assume the behaviour is repetitive for the whole year.

\begin{table}											
\centering 											
\begin{tabular}{|lcrrrr|}											
											
\hline 											
	&	Unit	&	Bldg1	&	Bldg2	&	Bldg3	&	Bldg4	\\
\hline 											
Demand,el	&	MWh/a	&	29.8	&	31.5	&	31.5	&	30.0	\\
Demand,th	&	MWh/a	&	113.0	&	100.9	&	58.0	&	40.4	\\
Occupants	&	\#	&	24	&	29	&	26	&	26	\\
BldgAge	&		&	$<1978$	&	1979-2001	&	$>2002$	&	$>2002$	\\
Insulation	&		&	standard	&	standard	&	standard	&	passive	\\
$A_{roof}$	&	$m^2$	&	176.0	&	166.8	&	125.6	&	125.6	\\
$A_{living}$	&	$m^2$	&	376.5	&	446.8	&	431.3	&	431.3	\\
											
\hline 											
\end{tabular}											
\caption{Input data of the four analyzed building types} 											
\label{tbl_inputdata_bldgtypes}											
\end{table}

\subsection{Component-wise and building-wise analyses}
\label{sec_setupcompwise}

The main analyses are divided into two sections. First, we study the different technological system components by themselves and in combination with each other for building 1. This so-called component-wise analysis allows us to identify the key economic drivers for a successful TEM. Table \ref{tbl_names_componentwise} presents the technological combinations and their abbreviations for this publication.

Secondly, the building-wise analysis studies the system \textit{COMBI} where the model can invest in all system components without considering EV. This analysis compares the heating and electricity profiles of four different building types, see Table \ref{tbl_inputdata_bldgtypes}. The first two building types represent the overwhelming majority of the MFH building stock. According to \cite{IWU.2015}, in terms of living area, this building type covers around $34\%$ and $26\%$ of the German building stock for MFHs, respectively. The last two building types represent a modern MFH with different insulation levels. Thus, the four buildings are presented in decreasing order of their heating demand.

\begin{table}																	
\centering 																	
																	
\begin{tabular}{|l|cccccccc|}																	
\hline																	
 &	\multicolumn{8}{c|}{Components}\\																
Name	&	Boi	&	HS	&	PV	&	Bat	&	HP	&	CHP	&	EV	&	EVopt	\\
\hline																	
REF	&	x	&	x	&		&		&		&		&		&		\\
PV	&	x	&	x	&	x	&		&		&		&		&		\\
PV\_BAT	&	x	&	x	&	x	&	x	&		&		&		&		\\
PV\_HP	&	x	&	x	&	x	&		&	x	&		&		&		\\
CHP	&	x	&	x	&		&		&		&	x	&		&		\\
CHP\_BAT	&	x	&	x	&		&	x	&		&	x	&		&		\\
CHP\_HP	&	x	&	x	&		&		&	x	&	x	&		&		\\
PV\_CHP	&	x	&	x	&	x	&		&		&	x	&		&		\\
PV\_CHP\_BAT	&	x	&	x	&	x	&	x	&		&	x	&		&		\\
COMBI	&	x	&	x	&	x	&	x	&	x	&	x	&		&		\\
COMBI\_EV	&	x	&	x	&	x	&	x	&	x	&	x	&	x	&		\\
COMBI\_EVopt	&	x	&	x	&	x	&	x	&	x	&	x	&		&	x	\\
\hline																	
																	
\multicolumn{9}{|l|}{\footnotesize{Boi: Gas boiler, HS: Heat storage, PV: Photovoltaic, Bat: Battery}}\\																	
\multicolumn{9}{|l|}{\footnotesize{HP: Heat pump, CHP: Combined heat and power}}\\																	
\multicolumn{9}{|l|}{\footnotesize{EV: 6 electric vehicles, EVopt: 6 Evs \& Optimized charging}}\\																	
\hline 																	
\end{tabular}																	
																	
\caption{System names of component wise analysis; \textit{x} marks which component option is included in the model run.} 																	
\label{tbl_names_componentwise}																	
																	
\end{table}

In both analyses, we apply a green-field approach, where the heating system in terms of generator and storage needs replacement, and no on-sight electricity generation exists. The reference case (\textit{REF}) for the economic performance of the TEM is the case where only a conventional gas boiler to cover the heating demand is installed, and electricity is drawn from the grid. The reference case is not participating in the TEL scheme. Fuel costs for the boiler are forwarded directly to the tenants, and the electricity is purchased by the tenants individually. For all other cases, we assume that all tenants participate in the TEM as the tenant electricity price is $10\%$ below the basic provider's tariff for an electricity demand between \kWh{2500} and \kWh{5000} per year \cite[p. 282]{Bundesnetzagentur.2020b}.

For this study, we alter $cap_{chp}$ in incremental steps of \kWel{10} up to \kWel{50}. A model formulation with a continuous variable for the CHP capacity results in very high run times, which conflicts with the study's goal to compare the TEM configurations. Thus, for every system of the component-wise analysis with a CHP unit as a technology option and every building in the building-wise analysis, we perform 6 model runs for $cap_{chp}$ between \kWel{0} and \kWel{50}. For the study's results, we use the model run and respective $cap_{chp}$ with the highest \textit{NPV}, which represents the objective value.

To solve the problem, we used the CPLEX solver and a maximum run time of 48 hours. The model is optimized on a Linux-based High-Performance Cluster with up to 150 GB RAM and 8 cores at 2.1 GHz in one single node. Up to 50 nodes in parallel carried out the various model runs.

\section{Results}
\label{sec_results}
This section divides the results into three parts. In the first Section\,\ref{section:results_comp}, we evaluate the systems design of the component-wise analysis and the influence of the design on the KPIs. After that, in Section\,\ref{section:results_bld}, we present the outcome for the building-wise analysis. Finally, we compare the influence of the most recent amendment of the TEL for January 2021 and the previous TEL in July 2020 on the NPV and the energy system layout in Section\,\ref{section:results_REL}.

\begin{figure}[h]
\includegraphics[width=\linewidth]{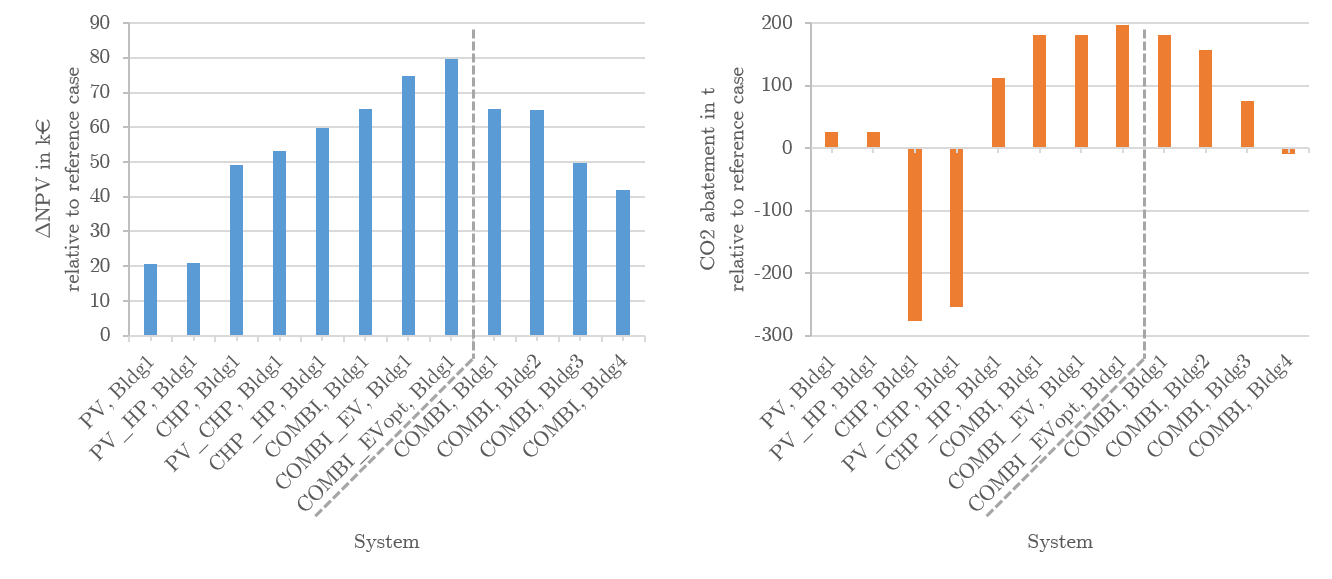}
\caption{Summary of \DeltaNPV{} and \CO{} abatement of the component and building-wise analysis. The line delimiter divides the two analyses.}
\label{fig.resultsNPVCO2}
\end{figure}

Figure \ref{fig.resultsNPVCO2} summarizes main results of Section\,\ref{section:results_comp} and Section\,\ref{section:results_bld}. It shows the \DeltaNPV{} and the \CO{} abatement of the different system designs. While the PV systems (\textit{PV} and \textit{PV\_HP}) have the lowest \DeltaNPV, the implementation of a CHP more than doubles the NPV. Moreover, this first glimpse indicates that technology diversity further increases the achievable NPV. A more detailed look reveals a complementary effect of the different technologies on the system's operation to utilize guaranteed subsidies more efficiently. Despite this complementarity, introducing a battery storage system was not found to increase the NPV\footnote{As part of a sensitivity analysis, we altered the variable part of the battery price in steps of \eurokWh{50}. The model only chose to install a battery for a variable price of \eurokWh{100} or less.}. That is why all the system results with a battery as an optional component are omitted in the following sections.
For the building-wise analysis, Figure \ref{fig.resultsNPVCO2} [right-hand sight of the delimiter] indicates a decrease of \DeltaNPV{} along with a reduction of the building's heating demand. Concerning the \CO{} abatement, Figure \ref{fig.resultsNPVCO2} shows mostly positive \CO{} abatement that correlates positively with \DeltaNPV. However, the CHP systems without an HP (\textit{CHP} and \textit{PV\_CHP}) as well as the \textit{COMBI} system for Bldg\,4 with the highest degree of insulation have negative \CO{} abatement. According to the polluter-pays principle, these systems emit more \CO{} than the reference case. However, these systems feed a large share of their self-generated electricity into the grid, which leads to relatively large quantities of \CO{} export compared to the other systems.

\subsection{Component-wise analysis}
\label{section:results_comp}

Table \ref{tbl_results_comp} summarizes the most relevant KPIs, the installed technologies, and capacities. 
The column \textit{REF} describes the reference case—the negative NPV results from accounting for the boiler's initial investment and its operational costs. The table depicts the unique systems designs found by the optimization in increasing order of \DeltaNPV{}\footnote{The result of the \textit{PV\_HP} system represents the \textit{PV} system's result as well. Both systems have the same design because in \textit{PV\_HP} no HP is chosen}. 
For system \textit{PV\_HP}, the PV system size of \kWp{10} complies with the first remuneration step of the TEL. Without any additional flexible electricity consumers such as an HP in the system, the self-consumption rate is $66.5\%$, and the remaining PV-electricity is fed into the grid. A conventional gas boiler covers the heat demand. Regarding the total \CO{} emissions, the abated amount of \CO{} emissions are relatively low with $13\%$ of the highest \CO{} abatement in system \textit{COMBI\_EVopt}. \DeltaCOexport{} with a value of \tCO{-11.3} indicates a negative amount of \CO{} emissions fed into the grid.

Comparing the \textit{PV\_HP} system and the \textit{CHP} system, \DeltaNPV{} more than doubles to a total of \kEUR{49.0}. The \textit{CHP} system, as well as the \textit{PV\_CHP} system, fully exploits the capacity limit of the CHP unit of \kWel{50} set by the TEL to reach the maximum NPV. Considering the full load hours, the CHP unit hardly operates outside of the subsidy scheme, while the model strictly restricts the \textit{PV\_CHP} system to the $30,000$ subsidised full load hours. Notably, in both systems, the CHP unit's peak power output is around \kWel{40} and thus less than the installed capacity. In both systems, the CHP unit covers most of the heat demand. The values of \SCR{} and \DA{} indicate that almost $80\%$ of the self-generated electricity is fed into the grid and the amount of self-generated electricity is more than 2.5 times as high as the electricity demand in the building. This leads to relatively high \CO{} emissions and consequently high negative values for \CO{} abatement according to the polluter pays principle. 
Large amounts of \CO{} emissions are exported, \tCO{380.2} and \tCO{381.4} respectively, which are greater than the negative value of the \CO{} abatement. However, the values of \DeltaCOexport{} with \tCO{125.8} and \tCO{120.4} respectively indicate that the feed-in increases the emission factor of grid electricity. 

Compared to the CHP only system, installing an HP (\textit{CHP\_HP}) increases \DeltaNPV{} by more than $20\%$ to around \kEUR{60}. The model chooses an HP with \kWth{12.7}. The HP introduces a flexible electricity demand to the system of roughly $60\%$. This demand is flexible in the sense that the HPs dispatch is an endogenous decision variable of the optimization model. Along with a relatively small CHP unit with \kWel{10}, the system achieves an \SCR{} of almost $100\%$, covering $89.4\%$ (\DSS{}) of the electricity demand by self-generated electricity. Together with a \DA{} of $89.6\%$, around $10\%$ of the electricity comes from the grid. This leads to abated \CO{} emissions of \tCO{111.4}. The full load hours of the CHP exceed the full load hour limit of the subsidies scheme by a factor of three. Hence, only one-third of the CHP hours are subsidized with the premium. This reduces the profit from electricity consumed by the tenants. Additionally, around $40\%$ of the CHP electricity is directed to the HP. 

In the case of total technological freedom (\textit{COMBI}), the model chooses to install PV, HP and CHP. Compared to the PV only system (\textit{PV}) and the CHP only system (\textit{CHP}), \DeltaNPV{} increases by more than $200\%$ and $33\%$ respectively. The PV system is sized greater than the \kWp{10} threshold of the first PV-remuneration step. Thus, this optimal system design can increase profits even though it has to compensate for a lower feed-in tariff and a lower SCP for PV electricity compared to the \textit{PV} system. This compensation is achieved by a higher self-consumption ($\mathit{SCR}=89.2\%$) as PV electricity is fed to the HP. $28\%$ of the PV electricity is directed to the HP and around $10\%$ to the grid. 

\begin{table}[!htbp]
    \resizebox{\linewidth}{!}{\begin{tabular}{|lcrrrrrrrr|}
\hline 
KPI & Unit &REF & PV\_ & CHP    & PV\_ & CHP\_ & COMBI & COMBI\_ & COMBI\_ \\ 
    &      &    & HP    &        & CHP   & HP     &       & EV       & EVopt    \\ 
\hline 
$NPV$ & \kEUR{} & -13.4 & 7.4 & 35.6 & 39.7 & 46.2 & 52.0 & 61.4 & 66.3 \\
$\Delta NPV$ & \kEUR{} & - & 20.8 & 49.0 & 53.1 & 59.7 & 65.4 & 74.8 & 79.8 \\
\hline 
\hline 
$D_{el,te}$ & MWh\slash a & 29.8 & 29.8 & 29.8 & 29.8 & 29.8 & 29.8 & 36.8 & 36.9 \\
$D_{el,hp}$ & MWh\slash a & - & - & - & - & 18.4 & 23.2 & 21.9 & 22.2 \\
$D_{el,tot}$ & MWh\slash a & 29.8 & 29.8 & 29.8 & 29.8 & 48.2 & 53.0 & 58.7 & 59.1 \\
\hline 
$cap_{PV}$ & kW & - & 10.0 & - & 7.6 & - & 12.6 & 15.0 & 15.0 \\
$cap_{CHP}$ & kWel & - & - & 50.0 & 50.0 & 10.0 & 20.0 & 20.0 & 20.0 \\
$cap_{HP}$ & kWth & - & - & - & - & 12.7 & 14.4 & 14.0 & 13.9 \\
$cap_{WWT}$ & kWh & - & - & 81.4 & 77.9 & 76.1 & 66.8 & 68.7 & 59.4 \\
$cap_{boil}$ & kWh & 60.3 & 60.3 & - & 1.0 & 12.4 & - & - & - \\
\hline 
$P_{chp,max}$ & kWel & - & - & 40.1 & 39.4 & 10.0 & 20.0 & 20.0 & 20.0 \\
$h_{chp,full}$ & hours & - & - & 30524 & 30000 & 86366 & 38745 & 41006 & 40239 \\
\hline 
$SCR_{el}$ & \% & - & 70.0 & 20.5 & 23.9 & 99.8 & 90.2 & 91.4 & 96.9 \\
$DSS_{el}$ & \% & - & 21.0 & 52.4 & 65.7 & 89.4 & 85.2 & 84.8 & 88.2 \\
$DA_{el}$ & \% & - & 30.0 & 256.0 & 274.5 & 89.6 & 94.4 & 92.8 & 91.0 \\
$GII$ & \% & - & 17.6 & 22.2 & 22.7 & 4.1 & 10.2 & 9.7 & 9.0 \\
$GII_{norm}$ & \% & - & 113.7 & 143.2 & 146.5 & 20.3 & 53.0 & 58.9 & 59.7 \\
\hline 
$CO_{2,ref}$ & t & 659.3 & 659.3 & 659.3 & 659.3 & 659.3 & 659.3 & 688.9 & 688.9 \\
$CO_{2,opt}$ & t & 659.3 & 633.1 & 935.9 & 913.4 & 548.0 & 477.9 & 508.5 & 491.4 \\
$\Delta CO_{2}$ & t & - & 26.2 & -276.6 & -254.1 & 111.4 & 181.4 & 180.4 & 197.5 \\
\hline 
$CO_{2,export}$ & t & - & - & 380.2 & 381.4 & 0.5 & 23.2 & 20.8 & 10.3 \\
$\Delta CO_{2,export}$ & t & - & -11.3 & 125.8 & 120.4 & 0.2 & 2.7 & 1.0 & 3.4 \\
\hline 
$CF_{subs}$ & \kEUR{} & - & 6.3 & 147.5 & 152.1 & 16.4 & 41.5 & 42.2 & 39.9 \\
$cac_{subs}$ & \eurotCO{} & - & 169.3 & - & - & 147.5 & 232.0 & 235.4 & 205.4 \\
\hline 
\end{tabular}
}
        \caption{Selected results for the component-wise analysis sorted by $\Delta NPV$ for an investment period of 20 years. Results for systems with a battery option are omitted, as no battery is installed.} 
    \label{tbl_results_comp}
\end{table}

Introducing the additional electricity demand of six EVs increases \DeltaNPV{} and the PV to \kWp{15}\footnote{It should be noted that the PV threshold of \kWp{15} is a result of internal PV remuneration steps of the optimization model, see Figure \ref{fig.feedinpremium}.} while simultaneously increasing the \SCR{} to $90.2\%$. Adding the option of optimized charging of these EVs, \DeltaNPV{} increases even further as the \SCR{} rises to $95.7\%$. The \DA{} decreases as the CHP generation is reduced, leading to the overall highest amount of \CO{} abatement. 
This high value for \CO{} abatement compared to the system \textit{COMBI} and \textit{COMBI\_EV} also leads to the lowest \CO{} abatement cost considering paid subsidies \cacsubs{}. Nonetheless, the lowest \cacsubs{} overall are attributed to the \textit{CHP\_HP} system which is followed by the \textit{PV\_HP} system. The high \SCR{} reveals that the system utilizes the high SCP for PV and CHP electricity. Additionally, as almost no CHP electricity is fed into the grid, almost all energy forms that account for \CO{} emissions are consumed within the MFB.
 
Regarding the grid interaction, GII and GII\tief{norm} enable a comparative analysis (Table\,\ref{tbl_results_comp}). 
High SCR, DSS and DA in combination with low GII values relate to systems that are favourable and highly beneficial from a grid system serving point of view. The systems with a high share of electricity fed into the grid (systems \textit{PV\_HP}, \textit{CHP}, and \textit{PV\_CHP}) demonstrate relatively high values for GII and GII\tief{norm}. This indicates high variations of electricity feed-in, which can be explained by the mismatch of the PV or CHP generation profile with the electricity demand profile. For the PV, these variations can be explained through the mismatch of irradiation and electricity demand profile (especially during summer), and for the CHP during the winter season with a mismatch between the electricity and heat demand profile and over-sizing of the CHP. The systems with an additional and flexible energy demand such as HP or load such as EVs present a lower profile variation or allow for higher capacities at similar levels of variations (e.g. PV capacity in COMBI vs. COMBI\tief{EV}).



\begin{figure}[h]
\centering
\includegraphics[width=\textwidth]{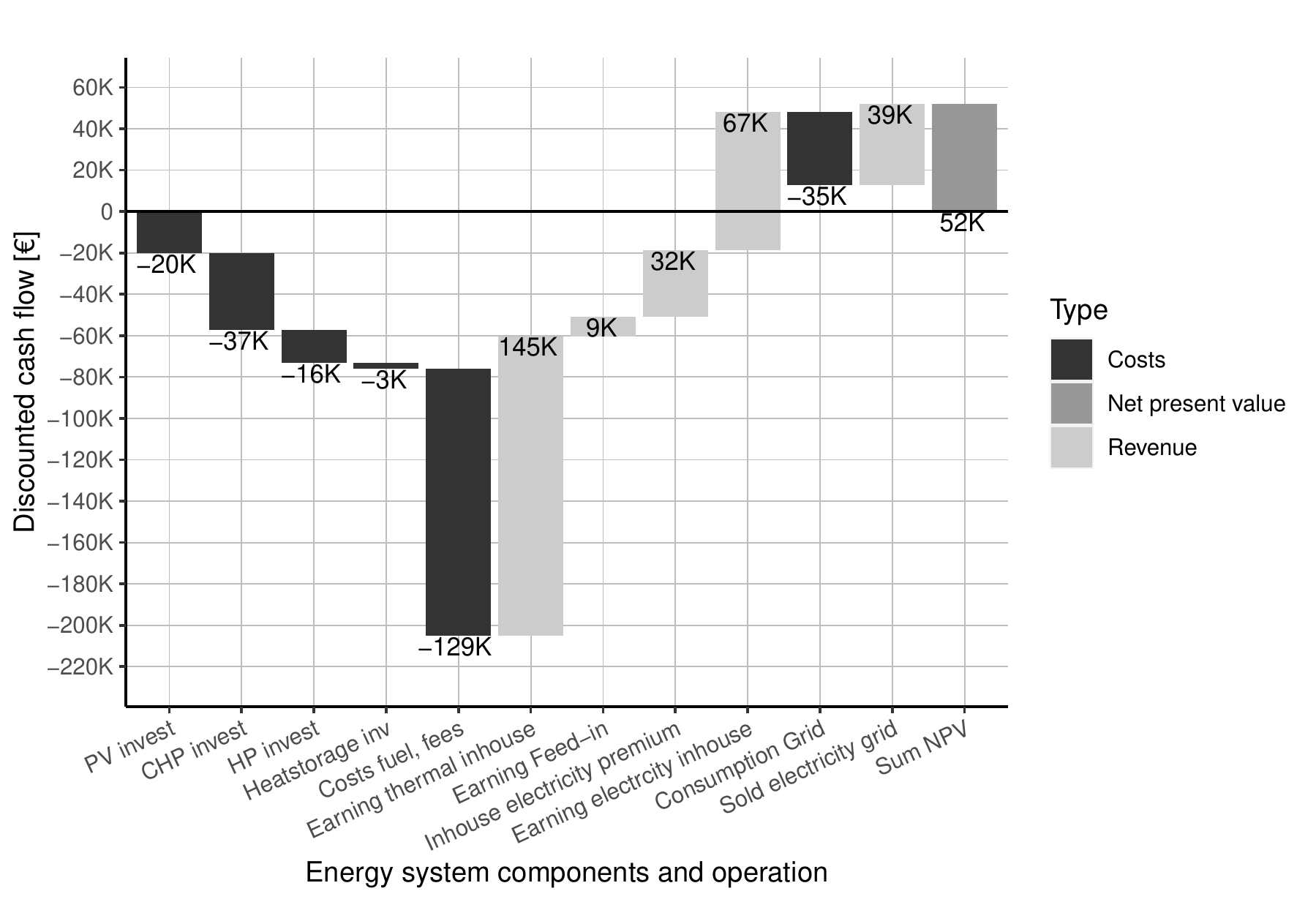}
\caption{Discounted cash flows and \textit{NPV} sorted by system components and operations of COMBI system in building 1.}
\label{fig.waterfall}
\end{figure}

Figure \ref{fig.waterfall} presents the waterfall diagram of discounted cash flows and the resulting \textit{NPV} for 20 years of operation, for system \textit{COMBI} for building 1. The largest investment of about \kEUR{36} is spent on the CHP unit. Operational costs for fuel to run the CHP unit and consumption fees make up the largest fraction of the negative cash flows. Despite a premium on electricity consumption, generation and feed-in, the returns from selling heat are a major component of compensating the initial and operational expenditures. This includes heat generated by the CHP as well as the HP. Concerning the other cash flows for self-generated electricity, the smallest part comes from feed-in tariffs followed by the revenue through the self-consumption premium. Selling self-generated electricity to the tenants is the second most substantial contribution that drives the NPV above zero. One peculiarity concerns the electricity drawn from the grid, whereby the landlord generates positive earnings. Due to the landlord's ability to bundle the electricity demand in one single contract, service providers offer lower prices to the landlord than the tenants for electricity from the grid. Therefore, slight earnings from grid consumption can be made. The large portion of cash flows accounting for self-consumed heat emphasizes that directing CHP electricity to the HP is a profitable option to increase the revenue from CHP electricity.

\begin{figure}[h]
\centering
\includegraphics[width=0.85\textwidth]{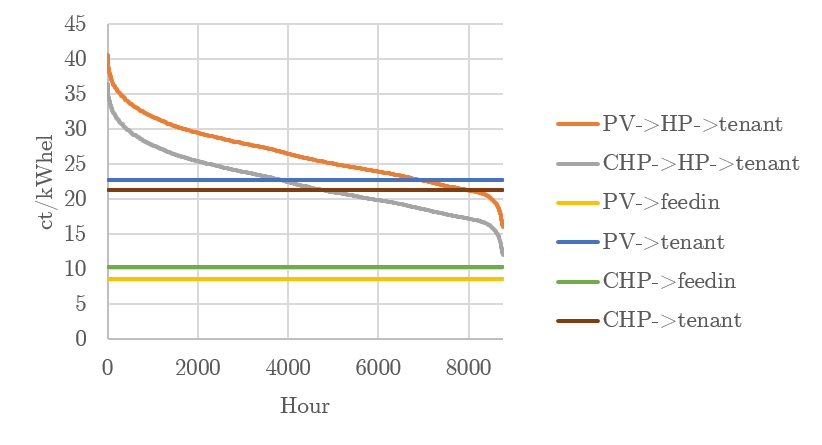}
\caption{Earnings from PV and CHP generated electricity as a sorted annual duration curve for the first year of operation; the earnings result from electricity direct self-consumption, feed-in or self-consumption in the HP which has an environmental-temperature-dependent COP.}
\label{fig.PV_CHP_HP_EnergyFlow}
\end{figure}

Figure \ref{fig.PV_CHP_HP_EnergyFlow} presents the sorted duration curve for the first year of operation for the earnings from directing self-generated electricity to the grid, the tenants, or converting it to heat via the HP and then selling it to the tenants. It illustrates the influence of COP on the earnings from HP self-consumption. Indeed for half the year, it is most profitable to directly self-consume PV- and CHP-generated electricity with the HP\footnote{In our study, the COP ranges from 1.6 in winter to almost 5 in summer. As long as the HP is operated with a COP above roughly 2.5 or 3, the earnings from converting the PV electricity or CHP electricity to heat surpass the earnings from selling the electricity directly to the tenants or feeding it into the grid.}. 
However, taking advantage of this is limited by the heat demand and respective storage capacities. Hence, the most profitable combination of PV-based HP heat supply is on the one hand limited by the size of the heat storage and the heat demand, whereby the latter is typically low when COP is highest. On the other hand, it is limited by the CHP operation, as heat is a by-product of the CHP operation and can only be consumed within the building. The cash flows and energy flows are further explained and illustrated in Figure\,\ref{App_InhouseCashFlow}.

\subsection{Results for different buildings types}
\label{section:results_bld}
Besides the analysis of technology combinations, the input parameters are varied in this section. We consider four different building types for the building-wise analysis, where the most characteristic features are the different total heat demands and demand profiles. Table \ref{tbl_results_bldgtypes} presents the building-wise analysis results in descending order of total annual heat demand. As to be expected, the model invests in smaller CHP units for the newer buildings, indicating that the heating demand bounds the size of the CHP. 
For building 1, 2, and 3, the model invests in an HP. Together with the HS, the HP provides a flexible energy demand. This allows for self-consumption rates of values greater than $88\%$. Considering the \DSS, the self-generated electricity satisfies between roughly $83\%$ and around $85\%$ of the electricity demand for all buildings. 

\begin{table}[h] 
\centering 
\begin{tabular}{|lcrrrr|}
\hline 
 	 & Unit & Bldg1 & Bldg2 & Bldg3 & Bldg4 \\ 
 	 & 	 & $<1978$ & 1979-2001 & $>2001$ & $>2001$passiv \\ \hline 
\hline 
$NPV$ & \kEUR{} & 52.0 & 50.4 & 38.2 & 33.0 \\
$\Delta NPV$ & \kEUR{} & 65.4 & 64.9 & 49.6 & 41.9 \\
\hline 
\hline 
$Q_{te}$ & MWh\slash a & 113.0 & 100.9 & 58.0 & 40.4 \\
$D_{el,te}$ & MWh\slash a & 29.8 & 31.5 & 31.5 & 30.0 \\
$D_{el,hp}$ & MWh\slash a & 23.2 & 19.6 & 7.8 & - \\
$D_{el,tot}$ & MWh\slash a & 53.0 & 51.1 & 39.3 & 30.0 \\
\hline 
$cap_{PV}$ & kW & 12.6 & 12.9 & 9.6 & 7.7 \\
$cap_{CHP}$ & kWel & 20.0 & 20.0 & 10.0 & 10.0 \\
$cap_{HP}$ & kWth & 14.4 & 12.8 & 6.2 & - \\
$cap_{WWT}$ & kWh & 66.8 & 71.9 & 78.4 & 53.3 \\
$cap_{boil}$ & kWh & - & - & 2.9 & 3.4 \\
\hline 
$P_{chp,max}$ & kWel & 20.0 & 20.0 & 10.0 & 10.0 \\
$h_{chp,full}$ & hours & 38745 & 36513 & 52473 & 49671 \\
\hline 
$SCR_{el}$ & \% & 90.2 & 89.6 & 95.4 & 78.9 \\
$DSS_{el}$ & \% & 85.2 & 84.3 & 84.5 & 83.4 \\
$DA_{el}$ & \% & 94.4 & 94.1 & 88.6 & 105.8 \\
$GII$ & \% & 10.2 & 10.5 & 9.1 & 12.7 \\
$GII_{norm}$ & \% & 53.0 & 57.3 & 56.7 & 89.4 \\
\hline 
$CO_{2,ref}$ & t & 659.3 & 609.0 & 406.6 & 316.7 \\
$CO_{2,opt}$ & t & 477.9 & 452.9 & 331.8 & 325.9 \\
$\Delta CO_{2}$ & t & 181.4 & 156.1 & 74.8 & -9.2 \\
\hline 
$CO_{2,export}$ & t & 23.2 & 23.6 & 3.2 & 33.1 \\
$\Delta CO_{2,export}$ & t & 2.7 & 2.7 & -3.6 & 5.0 \\
\hline 
$CF_{subs}$ & \kEUR{} & 41.5 & 41.7 & 21.4 & 26.5 \\
$cac_{subs}$ & \eurotCO{} & 232.0 & 272.2 & 272.8 & - \\
\hline 
\end{tabular}
\caption{Results for the 4 building types for system \textit{COMBI}} 
\label{tbl_results_bldgtypes}
\end{table}

Additionally, the results indicate that the heat demand restricts the size of the CHP unit and the HP as well. For building 3,  the unit size of the HP is reduced compared to building 1 and 2. However, in building 4, the model does not invest in an HP at all. With no HP, the system does not install any profitable solution to offer a flexible electricity demand. Consequently, the \SCR{} of $78\%$ is relatively low. Greater electricity feed-in leads to a \DA{} above $100\%$, which results in negative \CO{} abatement.
For building 2 and 3 \cacsubs{} are nearly identical and $17\%$ higher than for building 1. 
Regarding the grid interaction, multi-technology systems (PV, CHP, HP) seem to be NPV optimal with a feed-in variation corresponding to a GII of about $10\%$. This is especially noteworthy as demand profiles vary between the buildings.
Notably, compared to the other buildings, the level of \CO{} emissions in the reference case for building 4 is relatively low, which makes \CO{} abatement more challenging — building 1 with the highest reference \CO{} emissions indicates the highest \CO{} abatement as well as the lowest \cacsubs{}.

In Conclusion, a lower heat demand has a strong influence on the overall system design and performance. It reduces the capacity of the heat generators, like CHP and HP but of the PV system as well. Furthermore, this reduction influences the self-consumption potential and subsequently the profitability, as the NPV is lowest for building 4, \CO{} abatement and grid interaction.

\subsection{Comparison of the amendment from TEL 2020 to 2021}
\label{section:results_REL}

The TEL's most recent changes occurred in the REL between 2020 and 2021 and the CHPL between 2018 and 2020. Table \ref{tbl_changes2020_2021} gives an overview of these changes. Noticeable changes are the increase in the SCP for PV and CHP electricity and the increase of the feed-in tariff for CHP electricity. While the SCP for the CHP almost doubled, the feed-in tariff increased by around $37\%$. Simultaneously, the amendment reduces the maximum number of full load hours eligible for subsidies by one third.  

\begin{figure}[h]
\centering
\includegraphics[width=\textwidth]{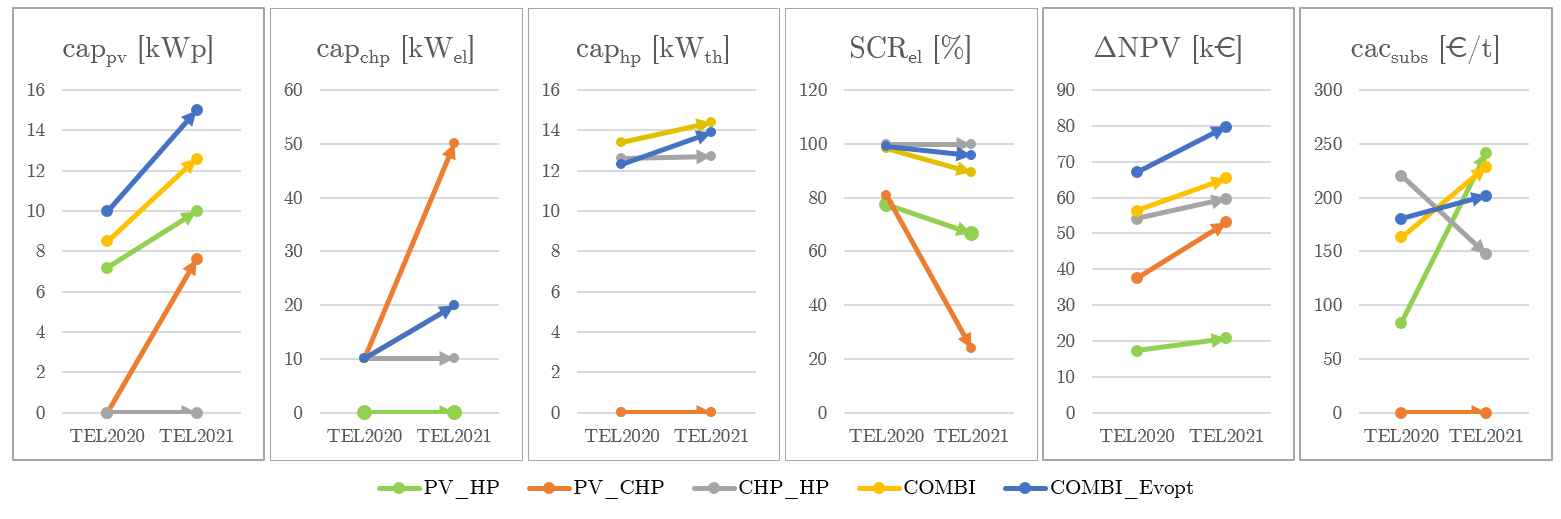}
\caption{Comparing the results of the tenant electricity model for the TEL in July 2020 and the TEL in January 2021 in building 1}
\label{fig.EEG2020_2021_compare_01}
\end{figure}

The model results for the implementation of the 2020 legislation are shown in Table \ref{tbl_results_comp_EEG2020}. Figure \ref{fig.EEG2020_2021_compare_01} illustrates the effect the legislative changes have on the model outcome, system setup and KPIs. The amendment of the law increases the installed PV capacity as well as the CHP capacity. In 2020, the model does not install a CHP unit larger than \kWel{10}. In 2021, the CHP capacities range between \kWel{10} and \kWel{50}. This can be interpreted as a result of the rise in CHP remuneration in combination with a reduction of the maximum full load hours. Generating electricity in the CHP unit yields higher profits than in 2020. Additionally, full load hours are defined as the ratio between total electricity output and installed capacity. Thus increasing the capacity allows the model to generate more electricity that falls under the TEL subsidy scheme. Eventually, the model exploits the provided limiting full load hour constraint.  A comparison of the results for system \textit{PV\_CHP} in 2020 and 2021 illustrates this effect. In 2020 the model invests in a \kWel{10} CHP unit and no PV. Considering the full load hours, almost $40\%$ of the CHP electricity does not receive subsidies and $80\%$ of the generated electricity is self-consumed. In 2021, the model installs the maximum CHP capacity of \kWel{50} and a PV system. The maximum full load hours are not exceeded and only $20\%$ of the electricity is self-consumed.

Overall, the systems in 2020 reach higher SCR values. At the same time, the \CO{} abatement and NPV in 2021 are more elevated. Nonetheless, along with the subsidies' increase from 2020 to 2021, the \COabat{} rise by a maximum factor of almost three. As a final notable observation, the results for system \textit{CHP\_HP} are relatively similar in both years. In this system, the HP offers high electric flexibility for the CHP dispatch. Both laws stimulate the system to reach a high SCR above $99.8\%$, and the same system setup achieves this.

\section{Discussion}
\label{sec_discussion}
In this section, we present our key findings and conclude policy implications. This is followed by a discussion about the uncertainties of the presented optimisation model and the resulting considerations for a real-life business application. Finally, we elaborate on future research topics that build upon this study. 

\subsection{Key findings and policy implications}

The component-wise analysis suggests that the introduction of additional and flexible electricity demand to the system increases the NPV of the tenant electricity model. This flexibility is the result of the additional shiftable electricity demand of an HP or a fleet of EVs. According to the TEM, self-consumption of electricity is always more profitable than feed-in. Therefore, an optimized HP dispatch or EV charging schedule increases the self-consumption rate and maximizes the NPV compared to a situation without them. Considering self-consumed PV electricity, flexibility through optimized EV-charging has a higher value than the HP dispatch due to the higher availability of EV-charging in summer and despite the favourable COP of the HP in summer.

Overall, the CHP operation can profit the most from the legal and subsidy framework, which renders the CHP unit a favourable technology for a profitable TEM. Furthermore, the results indicate that the heat demand has a strong influence on the investment and dispatch decisions for CHP and HP. From the building-wise analysis it is clear that larger building heat demands increase the NPV. More precisely, a greater heat demand allows for a larger CHP unit, the integration of an HP, and thus a high self-consumption rate. In the passive house with the lowest heat demand, investing in an HP is not profitable, which results in a low SCR, a relatively low NPV, and subsequently negative \CO{} abatement. In the latter case, additional electric flexibility like EV charging that is independent of the heating demand should be considered to increase the economic and ecological performance of the building.

This study shows that the TEM has the potential to reduce \CO{} emissions for most of the buildings analysed here. In fact for building 1, the system with the highest \CO{} abatement also presents the highest NPV for an investor. The amendment of the TEL in 2021 incentivizes increased PV investment and yields higher \CO{} abatement than the previous law in 2020. However, as the amendment in 2021 increases the amount of remuneration, the \COabcost{} also increases.

Nonetheless, the TEL is drafted strongly in favour of CHP-investment, which may be seen as a conflict with long-term \CO{} emission goals. In some cases, the reduction of the subsidized full load hours by the legislative changes leads to an over-sizing of the CHP-unit, which results in a large amount of grid feed-in and \CO{} emissions. The over-sizing could also impose additional stress on the grid as indicated by the GII. Furthermore, high insulation standards in buildings offer lower profitability for investors as the CHP's profitability increases with increasing heating demand. As building insulation standards continue to improve in future years, the economic case for such CHP systems will continue to diminish. 

Whilst PV operation by itself is profitable,more importantly, the TEL allows the investor to exploit the potential of combining different technology options.
Investing in a PV system by itself yields rather low profits, and integrating an HP into such a system is economically not feasible. 
However, while combining a CHP unit with an HP is already profitable, including a PV system yields even more benefits. It increases the NPV, and the HP allows for a larger PV system's profitable operation than in the PV only case resulting in greater \CO{} abatement. The PV system is even larger when considering optimised EV charging. This same system that combines all technologies yields the highest profit and \CO{} abatement. Thus, the combinatory technology options achieve larger PV instalments, which is the TEL's ultimate goal.

In conclusion and under the assumption of continuous decarbonisation of the power sector, the TEL can counteract \CO{} mitigation and energy conservation actions in a large share of buildings. Although CHPs are being considered as a bridging technology, in a 20 year time frame from now they may become a burden. Firstly, the grid decarbonisation can erode current fossil-based CHP benefits in efficiency and carbon emissions, and vice versa. Secondly, as heat demand strongly correlates with NPV, the incentives and arguments for energy conservation in old buildings fall out and a lock-in is generated. Investors would have to give up profits to save energy. Thirdly, less complex combinations e.g. CHP only or CHP-HP are more profitable than PV only or PV-CHP but do not contribute any additional renewable energies.
Hence, profit maximizing adoption with system layouts of low complexity can lead to lock-ins in Europe's largest residential building stock, which might slow down investments in renewable energies and heat demand reduction (energy conservation).

\subsection{Methodology and future research}
For this study, we assume a control and metering concept that allows for the proposed TEM and is already implemented in the buildings - other metering concepts, such as a totalizer considering the whole building and not the individual apartments, yield different results. On the tenants' side, we assume no changes in the number of tenants and their behaviour. Furthermore, we assume a 100\% participation in the TEM of all tenants over 20 years. On the landlord's side, we do not consider any additional financial burden, e.g. administrative measures, income or trade taxation considerations. Particularly the latter is a current hurdle for the adoption of tenant electricity.  Depending on the landlord's taxation situation, implementing a TEM can lead to income taxes or being taxed as a commercial entity. Moreover, inconclusive information on exemptions regarding trade taxes hinders the full consideration of all possible combinations\footnote{The exemption from trade taxes was announced together with the current REL amendment. Nonetheless, up until the 10th of February 2021 \cite{Bundestag.2021}, we could not identify any legal changes.}. These assumptions and simplifications need to be carefully considered, as they influence economic decisions.

In addition, one needs to consider the uncertainties as a result of the model formulation. We assume one representative year for the energy demand, driving, irradiation, and ambient temperature profiles over the investment period of 20 years. Additionally, the model operates with perfect foresight, which is a strong simplification for a real-life operation, especially for an energy storage system's optimal dispatch. The heat demand is also modelled without detailed consideration of heat inertia or return flow temperature. To allow for an adequate solving time, we selected discrete CHP capacities, see Section \ref{sec_studydesign}. All of these simplifications result in a significant uncertainty in the results presented, meaning they should be interpreted in this light. Nevertheless, whilst the absolute results are sensitive to variations in these input parameters, the overall trends in the results are felt to be more robust.

Furthermore, some of the analysed combinations achieve high degrees of electrical autonomy, which reduces the demand for grid electricity. This reduction would impact the contract between the landlord and grid service provider, resulting in an increasing electricity tariff. As well as influencing the tariff of a specific building, this newly-created energy community may also have wider impacts on the whole energy system. If a critical number of these communities become established, they may create a positive feedback loop \cite{McKenna.2018}, whereby increasing network fees distributed across fewer and fewer customers provide additional incentive for higher electrical autonomy \cite{Abada.2020}. We do not address this issue directly, as it is out of scope and requires a coupled and iterative approach with multiple system models. Nonetheless, Figure \ref{fig.waterfall} indicates this contractual uncertainty as a small risk.

The issue with over-sizing of CHP systems for MFBs may also suggest advantages in further aggregation. As well as economies of scale through larger plant sizes, aggregating to a district or neighbourhood level has the advantage of smoothing-out fluctuations in demand and supply \cite{McKenna.2017}. By also including some additional diversity of customers, optimizing an highly-renewable energy system at the district level can be economically and environmentally advantageous \cite{Orehounig.2015}. For example, it may ameliorate the encountered problem with large feed-ins of relatively high-\CO{}-intensity electricity. Appropriately dimensioned plants at the district level would have higher self-sufficiency rates and therefore a lower overall environmental impact. But the challenge of achieving high utilization rates of coupled heat production in summer still remains. In some contexts, the application of Seasonal Thermal Energy Storage (STES) systems may be appropriate, which would enable summer heat to be used in the winter months, but obviously has a significant influence on the costs \cite{McKenna.2019}.


Overall, then, our results show the complexity of the TEL and the possible adverse adoption based on incomplete or imperfect information. On the one hand, several simplifications and uncertainties indicate the limitations of the presented results. On the other hand, they emphasize the complexity for practitioners to implement a TEM in a real-life setting and give valuable insights into the different technologies' operational dependencies. It is therefore important to highlight the importance of the relationship between electricity and heating demand. Thus, future research should focus on renewables policy design to avoid adverse adoption and/or undesirable side-effects of subsidy schemes. In the present case, this means limiting the potential to oversize CHP units and feed large amounts of excess electricity into the grid. Furthermore, policy should adopt to an evolving energy system by closely considering and varying the EV driving profiles, charging infrastructure and other system-wide developments. Such analyses could be implemented through the application of model couplings between the sort of micro-level building model employed here and a whole energy system model at the national scale. Additionally, altering the tenants' structure via the number of tenants or the composition of age and social background affects the electricity and heating demand simultaneously \cite{McKenna.2016}. So this aspect should also be explored with linked models of socioeconomic technology adoption combined with spatially-disaggregated data on German households. Including insulation measures in the model would also be another step to depicting a more futuristic energy system, as only then could the competition effects between decentralised demand and supply sides be analysed. Furthermore, comparing different CHP modelling techniques or adding stochastic scenarios would help to better grasp the investment decision's complexity. In order to include some or all of the mentioned options while keeping the model solvable, the model complexity must be reduced. For example, an aggregated representation of the time series structure may be an option to achieve this complexity reduction. Finally, the results reveal non-intuitive coherence among the various energy and cash flows. Thus, investigating the implications of business models and legal frameworks in other countries could be of high interest for future work.

\section{Conclusion} 
\label{sec_conclusion}

In this paper, we formulate a MILP optimisation model to investigate the energy system design and operation of an MFB considering multiple technologies to match different electricity and heating demands. In contrast to other publications, this model includes multiple technological options such as PV, CHP, HP, and EV charging. The novelty is the application to the German case of the Tenant Electricity Law (TEL), a framework for local energy communities. We determine the optimal system design for different technology combinations and building types. Additionally, we compare the effect of the TEL amendment from 2020 to the current version in 2021. We analyse the economic and ecological performance as well as the interaction with the national grid. 

This comprehensive model and the study of the legal framework, technological variety and the principle agent distinction discloses the merits and pitfalls of energy communities. The key findings show that the latest amendment of the TEL in Germany increases the profitability of a TEM in MFBs. Hence, the amendment possibly increases the attractiveness of its adoption and accelerates its diffusion in MFHs, but the current framework favours the CHP technology over other options. Nonetheless, the law fosters the various technologies' dependencies as the highest profitability is achieved when combining PV, CHP, HP and EV. However, the current TEL may fail to support national \CO{} mitigation goals. First, this is because the TEM's profitability depends on the heat demand and a reduction of the heat demand, through insulation measures, might coincide with a profitability reduction. Secondly, it needs to be discussed if subsidising CHP operation as a bridging technology achieves the desired levels of \CO{} abatement.

Tapping renewable energy sources and increasing energy conservation in the built environment through energy communities seems to be a model of choice in many countries to reach climate mitigation goals.
Hence this study contributes additional insights to the international scientific and policy discussion around energy communities. The implementation of energy communities differs greatly by country, with forms of the TEL in the UK (Private Wire Networks policy), Spain (Collective Auto Consumption policy), Netherlands (Post Code Rose policy). By adding another case study application to this research, this paper has further extended the the existing knowledge about cost- and environmentally-effective applications to also include the current German situation. Furthermore, the gain of insight through the developed detailed approach provides a persuasive precedent for international practitioners, policymakers and investigating other energy community frameworks in detail.

\appendix
\section{Additional tables}

\begin{table}											
\centering 											
\begin{threeparttable}											
\begin{tabular}{|llll|}											
\hline 											
parameter	&	description	&	Unit	&	Value	\\ \hline				
$c_{REL}$	&	REL levy	&	\eurokWh{}	&	0.0650	\\				
VAT	&	Value added taxes	&	\%	&	19	\\				
$c_{M\&I}$	&	cost metering \& invoicing	&	\eurokWh{}	&	0.0061	\\				
$c_{chp,te}$\tnote{1}	&	CHP self-consumption premium	&	\eurokWh{}	&	0.0800	\\				
$c_{chp,feedin}$\tnote{1}	&	CHP feed in	&	\eurokWh{}	&	0.1600	\\				
$c_{inv,fix}^{PV,a=0}$	&	PV, fix cost	&	\euro{}	&	0	\\				
$c_{inv,var}^{PV,a=0}$	&	PV, variable cost	&	\eurokWp{}	&	1194.39	\\				
$c_{inv,fix}^{chp,a=0}$	&	CHP, fix cost	&	\euro{}	&	15000	\\				
$c_{inv,var}^{chp,a=0}$	&	CHP, variable cost	&	\eurokWel{}	&	970.30	\\				
$c_{inv,fix}^{bat,a=0}$	&	Battery, fix cost	&	\euro{}	&	2000	\\				
$c_{inv,var}^{bat,a=0}$	&	Battery, variable cost	&	\eurokWh{}	&	530.84	\\				
$c_{inv,fix}^{inv,a=0}$	&	Inverter, fix cost	&	\euro{}	&	0	\\				
$c_{inv,var}^{inv,a=0}$	&	Inverter, variable cost	&	\eurokWel{}	&	250	\\				
$c_{inv,fix}^{hp,a=0}$	&	Heat pump, fix cost	&	\euro{}	&	5000	\\				
$c_{inv,var}^{hp,a=0}$	&	Heat Pump, variable cost	&	\eurokWth{}	&	582	\\				
$c_{inv,fix}^{boiler}$	&	Boiler, fix cost	&	\euro{}	&	0	\\				
$c_{inv,var}^{boiler}$	&	Boiler, variable cost	&	\eurokWth{}	&	175	\\ \hline 				
											
\end{tabular}											
\begin{tablenotes}											
\item[1] $cap^{CHP}<50kWel$											
\end{tablenotes}											
\end{threeparttable}											
\caption{Input data 1} 											
\label{tbl_inputdata1}											
\end{table}											
											
\begin{table}											
\centering											
\begin{tabular}{|llllll|}											
\hline 											
parameter	&	Unit	&	Value	&	parameter	&	Unit	&	Value	\\
\hline 											
$A$	&	$\mathit{years}$	&	20	&	$\eta_{chp,el}$	&	 -	&	0.35	\\
$T$	&	$\mathit{hourd}$	&	8760	&	$\eta_{chp,th}$	&	 -	&	0.58	\\
$i$	&	$\%$	&	4	&	$\sigma_{chp}$	&	 -	&	0.60	\\
$r_{el}$	&	$\%$	&	2	&	$r_{chp,min}$	&	$\%$	&	40	\\
$EF_{el,grid,2019}$	&	$g_{CO2}\slash kWh_{el}$	&	401	&	$cap_{lim,REL}$	&	\kWel{}	&	10	\\
$r_{EFgrid}$	&	$\%$	&	-6	&	$E_{chp,lim,REL}$	&	\MWh{}	&	10	\\
$EF_{gas,2019}$	&	$g_{CO2}\slash kWh$	&	201	&	$EF_{chp,el}$	&	$g_{CO2}\slash kWh_{el}$	&	313	\\
$r_{EFgas}$	&	$\%$	&	0	&	$\eta_{el,altl}$	&	 -	&	0.40	\\
$EF_{el,PV,2019}$	&	$g_{CO2}\slash kWh_{el}$	&	0	&	$\eta_{th,alt}$	&	 -	&	0.90	\\
$r_{EF,PV}$	&	$\%$	&	-6	&		&		&		\\
											
\hline 											
\end{tabular}											
\caption{Input data 2} 											
\label{tbl_inputdata2}											
											
\end{table}

\begin{table}
\centering 									
\begin{tabular}{|lcccc|}									
\hline 									
Year	&	$c_{el,landlord}$	&	$c_{el,tenant}$	&	$c_{gas}$	&	$c_{CO2}$	\\ \hline
	&	\eurokWh{}	&	\eurokWh{}	&	\eurokWh{}	&	\eurotCO{}	\\
2021	&	0.2973	&	0.3293	&	0.0633	&	25	\\
2022	&	0.3033	&	0.3359	&	0.0654	&	30	\\
2023	&	0.3094	&	0.3426	&	0.0676	&	35	\\
2024	&	0.3155	&	0.3494	&	0.0709	&	45	\\
2025	&	0.3219	&	0.3564	&	0.0741	&	55	\\
2026	&	0.3283	&	0.3635	&	0.0754	&	55	\\
2027	&	0.3349	&	0.3708	&	0.0766	&	55	\\
2028	&	0.3416	&	0.3782	&	0.0780	&	55	\\
2029	&	0.3484	&	0.3858	&	0.0793	&	55	\\
2030	&	0.3554	&	0.3935	&	0.0827	&	65	\\
2031	&	0.3625	&	0.4014	&	0.0841	&	65	\\
2032	&	0.3697	&	0.4094	&	0.0855	&	65	\\
2033	&	0.3771	&	0.4176	&	0.0869	&	65	\\
2034	&	0.3846	&	0.4260	&	0.0884	&	65	\\
2035	&	0.3923	&	0.4345	&	0.0919	&	75	\\
2036	&	0.4002	&	0.4432	&	0.0935	&	75	\\
2037	&	0.4082	&	0.4520	&	0.0950	&	75	\\
2038	&	0.4164	&	0.4611	&	0.0966	&	75	\\
2039	&	0.4247	&	0.4703	&	0.0983	&	75	\\
2040	&	0.4332	&	0.4797	&	0.0999	&	75	\\ \hline 
									
\end{tabular}									

\caption{Consumer prices for landlord and tenants} 
\label{tbl_consumerprices}

\end{table}

\begin{table}									
\centering 									
\begin{tabular}{|ccccc|}									
\hline 									
									
\textit{rs}	&	$cap_{pv}^{rs}$	&	$c_{pv,scp}$ 	&	$c_{pv,feedin}$ 	&	$c_{pv,levy}$ 	\\
Remuneration scheme	&	Upper limit	&	\eurokWh{}	&	\eurokWh{}	&	\eurokWh{}	\\
1	&	10	&	0.0379	&	0.0856	&	-	\\
2	&	15	&	0.0374	&	0.0851	&	-	\\
3	&	20	&	0.0367	&	0.0846	&	-	\\
4	&	25	&	0.0364	&	0.0843	&	-	\\
5	&	30	&	0.0362	&	0.0841	&	-	\\
6	&	35	&	0.0360	&	0.0840	&	0.4351	\\
7	&	40	&	0.0359	&	0.0839	&	0.4351	\\
8	&	45	&	0.0352	&	0.0828	&	0.4351	\\
9	&	50	&	0.0340	&	0.0811	&	0.4351	\\
10	&	55	&	0.0330	&	0.0797	&	0.4351	\\
11	&	60	&	0.0322	&	0.0785	&	0.4351	\\
12	&	65	&	0.0315	&	0.0775	&	0.4351	\\
13	&	70	&	0.0309	&	0.0767	&	0.4351	\\
14	&	75	&	0.0304	&	0.0760	&	0.4351	\\
15	&	80	&	0.0300	&	0.0753	&	0.4351	\\
16	&	85	&	0.0296	&	0.0748	&	0.4351	\\
17	&	90	&	0.0293	&	0.0743	&	0.4351	\\
18	&	95	&	0.0290	&	0.0738	&	0.4351	\\
19	&	100	&	0.0287	&	0.0735	&	0.4351	\\
									
\hline 									
\end{tabular}									
\caption{Tenant self-consumption premium, feed-in tariff and self-consumption levy for PV-electricity depending on the remuneration scheme and installed PV capacity respectively.} 									
\label{tbl_PV_remuneration}									
\end{table}

\begin{table}									
\centering 									
\begin{tabular}{|lllrr|}									
\hline 									
									
Parameter	&	Capacity limit	&	Unit	&	01.07.2020	&	01.01.2021	\\
\hline									
\multirow{3}{*}{$\mathit{SCP_{pv}}$}	&	$\mathit{cap_{pv}}<= 10\mathit{kWp}$	&	\ctkWh{}	&	0.53	&	3.79	\\
	&	$\mathit{cap_{pv}}<= 40\mathit{kWp}$	&	\ctkWh{}	&	0.28	&	3.52	\\
	&	$\mathit{cap_{pv}}<= 750\mathit{kWp}$	&	\ctkWh{}	&	-1.11	&	2.37	\\
\hline									
\multirow{3}{*}{$\mathit{feed-in_{pv}}$}	&	$\mathit{cap_{pv}}<= 10\mathit{kWp}$	&	\ctkWh{}	&	9.03	&	8.56	\\
	&	$\mathit{cap_{pv}}<= 40\mathit{kWp}$	&	\ctkWh{}	&	8.78	&	8.33	\\
	&	$\mathit{cap_{pv}}<= 750\mathit{kWp}$	&	\ctkWh{}	&	6.89	&	6.62	\\
\hline									
$\mathit{SCP_{chp}}$	&		&	\ctkWh{}	&	4.10	&	8.00	\\
$\mathit{feed-in_{chp}}$	&		&	\ctkWh{}	&	11.66	&	16.00	\\
$\mathit{h_{chp,fullload}}$	&		&	hours	&	45000	&	30000	\\
\hline									
		
\end{tabular}									
\caption{Most important changes between legislation on 01.07.2020 and 01.01.2021.} 									
\label{tbl_changes2020_2021}									
\end{table}

\begin{table}[!htbp]
    \resizebox{\linewidth}{!}{\begin{tabular}{|lcrrrrrrrr|}
\hline 
KPI & Unit &REF & PV\_ & PV\_ & CHP    & CHP\_ & COMBI & COMBI\_ & COMBI\_ \\ 
    &      &    & HP    & CHP   &        & HP     &       & EV       & EVopt    \\ 
\hline 
$NPV$ & \kEUR{} & -13.4 & 3.9 & 24.0 & 24.0 & 40.7 & 43.0 & 50.0 & 53.7 \\
$\Delta NPV$ & \kEUR{} & - & 17.3 & 37.4 & 37.5 & 54.1 & 56.4 & 63.5 & 67.1 \\
\hline 
\hline 
$D_{el,te}$ & MWh\slash a & 29.8 & 29.8 & 29.8 & 29.8 & 29.8 & 29.8 & 36.8 & 36.9 \\
$D_{el,hp}$ & MWh\slash a & - & - & - & - & 18.4 & 20.6 & 18.7 & 18.3 \\
$D_{el,tot}$ & MWh\slash a & 29.8 & 29.8 & 29.8 & 29.8 & 48.2 & 50.4 & 55.6 & 55.1 \\
\hline 
$cap_{PV}$ & kW & - & 7.2 & - & - & - & 8.5 & 10.0 & 10.0 \\
$cap_{CHP}$ & kWel & - & - & 10.0 & 10.0 & 10.0 & 10.0 & 10.0 & 10.0 \\
$cap_{HP}$ & kWth & - & - & - & - & 12.6 & 13.4 & 13.1 & 12.3 \\
$cap_{WWT}$ & kWh & - & - & 63.8 & 63.9 & 76.6 & 75.9 & 76.2 & 73.9 \\
$cap_{boil}$ & kWh & 60.3 & 60.3 & 25.7 & 25.7 & 12.4 & 11.8 & 12.0 & 13.1 \\
\hline 
$P_{chp,max}$ & kWel & - & - & 10.0 & 10.0 & 10.0 & 10.0 & 10.0 & 10.0 \\
$h_{chp,full}$ & hours & - & - & 62436 & 62582 & 86214 & 78101 & 82484 & 87342 \\
\hline 
$SCR_{el}$ & \% & - & 81.5 & 80.7 & 80.6 & 99.9 & 99.0 & 98.8 & 100.0 \\
$DSS_{el}$ & \% & - & 17.8 & 84.5 & 84.6 & 89.3 & 91.8 & 89.4 & 95.7 \\
$DA_{el}$ & \% & - & 21.8 & 104.7 & 105.0 & 89.4 & 92.7 & 90.5 & 95.7 \\
$GII$ & \% & - & 14.1 & 13.2 & 13.2 & 3.3 & 5.4 & 5.8 & 2.2 \\
$GII_{norm}$ & \% & - & 91.1 & 85.0 & 85.1 & 16.1 & 27.1 & 46.8 & 12.2 \\
\hline 
$CO_{2,ref}$ & t & 667.3 & 667.3 & 667.3 & 667.3 & 667.3 & 667.3 & 698.7 & 698.7 \\
$CO_{2,opt}$ & t & 667.3 & 643.7 & 695.4 & 695.6 & 549.3 & 496.6 & 535.6 & 538.2 \\
$\Delta CO_{2}$ & t & - & 23.6 & -28.1 & -28.3 & 118.0 & 170.8 & 163.1 & 160.6 \\
\hline 
$CO_{2,export}$ & t & - & - & 37.8 & 38.0 & 0.4 & 1.3 & 1.4 & - \\
$\Delta CO_{2,export}$ & t & - & -5.4 & 10.9 & 10.9 & 0.1 & -0.8 & -1.3 & - \\
\hline 
$CF_{subs}$ & \kEUR{} & - & 1.9 & 18.7 & 18.8 & 12.6 & 13.3 & 13.6 & 12.9 \\
$cac_{subs}$ & \eurotCO{} & - & 64.0 & - & - & 106.8 & 77.5 & 82.6 & 80.6 \\
\hline 
\end{tabular}
}
        \caption{Results for component wise analysis sorted by $\Delta NPV$ for comparison of the TEL in July 2020 and the TEL in January 2021} 
    \label{tbl_results_comp_EEG2020}
\end{table}
\newpage

\bibliographystyle{elsarticle-num-names}
\bibliography{Tenant_electricity_model.bib}

\begin{thebibliography}{57}
\expandafter\ifx\csname natexlab\endcsname\relax\def\natexlab#1{#1}\fi
\providecommand{\url}[1]{\texttt{#1}}
\providecommand{\href}[2]{#2}
\providecommand{\path}[1]{#1}
\providecommand{\DOIprefix}{doi:}
\providecommand{\ArXivprefix}{arXiv:}
\providecommand{\URLprefix}{URL: }
\providecommand{\Pubmedprefix}{pmid:}
\providecommand{\doi}[1]{\href{http://dx.doi.org/#1}{\path{#1}}}
\providecommand{\Pubmed}[1]{\href{pmid:#1}{\path{#1}}}
\providecommand{\bibinfo}[2]{#2}
\ifx\xfnm\relax \def\xfnm[#1]{\unskip,\space#1}\fi
\bibitem[{{European Parliament and the Council}(2018)}]{EU.2018}
\bibinfo{author}{{European Parliament and the Council}},
\newblock \bibinfo{title}{amending directive 2010/31/eu on the energy
  performance of buildings and directive 2012/27/eu on energy efficiency}
  (\bibinfo{year}{2018}). \URLprefix
  \url{https://eur-lex.europa.eu/legal-content/EN/TXT/PDF/?uri=CELEX:02010L0031-20181224&from=EN}.
\bibitem[{{European Commission}(2018)}]{EU.2018b}
\bibinfo{author}{{European Commission}}, \bibinfo{title}{Directive on the
  promotion of the use of energy from renewable sources (recast): Eu 2018/
  2001}, \bibinfo{year}{2018}. \URLprefix
  \url{https://eur-lex.europa.eu/legal-content/EN/TXT/PDF/?uri=CELEX:32018L2001&amp;from=en}.
\bibitem[{{European Parliament and the Council of the EU}(2019)}]{EU.2019}
\bibinfo{author}{{European Parliament and the Council of the EU}},
  \bibinfo{title}{Directive on common rules for the internal market for
  electricity and amending: Eu 2019/944}, \bibinfo{year}{2019}. \URLprefix
  \url{https://eur-lex.europa.eu/legal-content/EN/TXT/PDF/?uri=CELEX:32019L0944&amp;from=EN}.
\bibitem[{In{\^e}s et~al.(2020)In{\^e}s, Guilherme, Esther, Swantje, Stephen,
  and Lars}]{Ines.2020}
\bibinfo{author}{C.~In{\^e}s}, \bibinfo{author}{P.~L. Guilherme},
  \bibinfo{author}{M.-G. Esther}, \bibinfo{author}{G.~Swantje},
  \bibinfo{author}{H.~Stephen}, \bibinfo{author}{H.~Lars},
\newblock \bibinfo{title}{Regulatory challenges and opportunities for
  collective renewable energy prosumers in the eu},
\newblock \bibinfo{journal}{Energy Policy} \bibinfo{volume}{138}
  (\bibinfo{year}{2020}) \bibinfo{pages}{111212}.
  \DOIprefix\doi{10.1016/j.enpol.2019.111212}.
\bibitem[{{F.G. Reis} et~al.(2021){F.G. Reis}, Gon{\c{c}}alves, {A.R. Lopes},
  and {Henggeler Antunes}}]{Reis.2021}
\bibinfo{author}{I.~{F.G. Reis}}, \bibinfo{author}{I.~Gon{\c{c}}alves},
  \bibinfo{author}{M.~{A.R. Lopes}}, \bibinfo{author}{C.~{Henggeler Antunes}},
\newblock \bibinfo{title}{Business models for energy communities: A review of
  key issues and trends},
\newblock \bibinfo{journal}{Renewable and Sustainable Energy Reviews}
  \bibinfo{volume}{144} (\bibinfo{year}{2021}) \bibinfo{pages}{111013}.
  \DOIprefix\doi{10.1016/j.rser.2021.111013}.
\bibitem[{Abada et~al.(2020)Abada, Ehrenmann, and Lambin}]{Abada.2020}
\bibinfo{author}{I.~Abada}, \bibinfo{author}{A.~Ehrenmann},
  \bibinfo{author}{X.~Lambin},
\newblock \bibinfo{title}{Unintended consequences: The snowball effect of
  energy communities},
\newblock \bibinfo{journal}{Energy Policy} \bibinfo{volume}{143}
  (\bibinfo{year}{2020}) \bibinfo{pages}{111597}.
  \DOIprefix\doi{10.1016/j.enpol.2020.111597}.
\bibitem[{{Bundesministerium f{\"u}r Umwelt}(2020)}]{BMU.2020}
\bibinfo{author}{{Bundesministerium f{\"u}r Umwelt}},
  \bibinfo{title}{Klimaschutzbericht 2019: zum aktionsprogramm klimaschutz 2020
  der bundesregierung}, \bibinfo{year}{2020}. \URLprefix
  \url{https://www.bmu.de/fileadmin/Daten_BMU/Download_PDF/Klimaschutz/klimaschutzbericht_2019_kabinettsfassung_bf.pdf}.
\bibitem[{Bigalke et~al.(2016)Bigalke, Kunde, Schmitt, Zeng, Discher, Bensmann,
  and Stolte}]{Bigalke.2016}
\bibinfo{author}{U.~Bigalke}, \bibinfo{author}{J.~Kunde},
  \bibinfo{author}{M.~Schmitt}, \bibinfo{author}{Y.~Zeng},
  \bibinfo{author}{H.~Discher}, \bibinfo{author}{K.~Bensmann},
  \bibinfo{author}{C.~Stolte}, \bibinfo{title}{Der dena-geb{\"a}udereport
  2016}, \bibinfo{year}{2016}, \bibinfo{editor}{{Deutsche Energie-Agentur
  GmbH}} (Ed.).
\bibitem[{{Stephen Hall} et~al.(2020){Stephen Hall}, {Donal Brown}, {Mark
  Davis}, Ehrtmann, and Holstenkamp}]{Hall.2020}
\bibinfo{author}{{Stephen Hall}}, \bibinfo{author}{{Donal Brown}},
  \bibinfo{author}{{Mark Davis}}, \bibinfo{author}{M.~Ehrtmann},
  \bibinfo{author}{L.~Holstenkamp},
\newblock \bibinfo{title}{Business models for prosumers in europe. proseu -
  prosumers for the energy union: Mainstreaming active participation of
  citizens in the energy transition: (deliverable nod4.1)}
  (\bibinfo{year}{2020}).
\bibitem[{Mancarella(2014)}]{Mancarella.2014}
\bibinfo{author}{P.~Mancarella},
\newblock \bibinfo{title}{Mes (multi-energy systems): An overview of concepts
  and evaluation models},
\newblock \bibinfo{journal}{Energy} \bibinfo{volume}{65} (\bibinfo{year}{2014})
  \bibinfo{pages}{1--17}. \DOIprefix\doi{10.1016/j.energy.2013.10.041}.
\bibitem[{Heendeniya et~al.(2020)Heendeniya, Sumper, and
  Eicker}]{Heendeniya.2020}
\bibinfo{author}{C.~B. Heendeniya}, \bibinfo{author}{A.~Sumper},
  \bibinfo{author}{U.~Eicker},
\newblock \bibinfo{title}{The multi-energy system co-planning of nearly
  zero-energy districts -- status-quo and future research potential},
\newblock \bibinfo{journal}{Applied Energy} \bibinfo{volume}{267}
  (\bibinfo{year}{2020}) \bibinfo{pages}{114953}.
  \DOIprefix\doi{10.1016/j.apenergy.2020.114953}.
\bibitem[{Lindberg et~al.(2016)Lindberg, Fischer, Doorman, Korp{\aa}s, and
  Sartori}]{Lindberg.2016}
\bibinfo{author}{K.~B. Lindberg}, \bibinfo{author}{D.~Fischer},
  \bibinfo{author}{G.~Doorman}, \bibinfo{author}{M.~Korp{\aa}s},
  \bibinfo{author}{I.~Sartori},
\newblock \bibinfo{title}{Cost-optimal energy system design in zero energy
  buildings with resulting grid impact: A case study of a german multi-family
  house},
\newblock \bibinfo{journal}{Energy and Buildings} \bibinfo{volume}{127}
  (\bibinfo{year}{2016}) \bibinfo{pages}{830--845}.
  \DOIprefix\doi{10.1016/j.enbuild.2016.05.063}.
\bibitem[{Fina et~al.(2019)Fina, Auer, and Friedl}]{Fina.2019}
\bibinfo{author}{B.~Fina}, \bibinfo{author}{H.~Auer},
  \bibinfo{author}{W.~Friedl},
\newblock \bibinfo{title}{Profitability of active retrofitting of
  multi-apartment buildings: Building-attached/integrated photovoltaics with
  special consideration of different heating systems},
\newblock \bibinfo{journal}{Energy and Buildings} \bibinfo{volume}{190}
  (\bibinfo{year}{2019}) \bibinfo{pages}{86--102}.
  \DOIprefix\doi{10.1016/j.enbuild.2019.02.034}.
\bibitem[{Fina et~al.(2018)Fina, Fleischhacker, Auer, and Lettner}]{Fina.2018}
\bibinfo{author}{B.~Fina}, \bibinfo{author}{A.~Fleischhacker},
  \bibinfo{author}{H.~Auer}, \bibinfo{author}{G.~Lettner},
\newblock \bibinfo{title}{Economic assessment and business models of rooftop
  photovoltaic systems in multiapartment buildings: Case studies for austria
  and germany},
\newblock \bibinfo{journal}{Journal of Renewable Energy} \bibinfo{volume}{2018}
  (\bibinfo{year}{2018}) \bibinfo{pages}{1--16}.
  \DOIprefix\doi{10.1155/2018/9759680}.
\bibitem[{Ferrara et~al.(2018)Ferrara, Sirombo, and Fabrizio}]{Ferrara.2018}
\bibinfo{author}{M.~Ferrara}, \bibinfo{author}{E.~Sirombo},
  \bibinfo{author}{E.~Fabrizio},
\newblock \bibinfo{title}{{Energy-optimized versus cost-optimized design of
  high-performing dwellings: The case of multifamily buildings}},
\newblock \bibinfo{journal}{Science and Technology for the Built Environment}
  \bibinfo{volume}{24} (\bibinfo{year}{2018}) \bibinfo{pages}{513--528}.
  \DOIprefix\doi{10.1080/23744731.2018.1448656}.
\bibitem[{Comodi et~al.(2015)Comodi, Giantomassi, Severini, Squartini,
  Ferracuti, Fonti, {Nardi Cesarini}, Morodo, and Polonara}]{COMODI2015854}
\bibinfo{author}{G.~Comodi}, \bibinfo{author}{A.~Giantomassi},
  \bibinfo{author}{M.~Severini}, \bibinfo{author}{S.~Squartini},
  \bibinfo{author}{F.~Ferracuti}, \bibinfo{author}{A.~Fonti},
  \bibinfo{author}{D.~{Nardi Cesarini}}, \bibinfo{author}{M.~Morodo},
  \bibinfo{author}{F.~Polonara},
\newblock \bibinfo{title}{Multi-apartment residential microgrid with electrical
  and thermal storage devices: Experimental analysis and simulation of energy
  management strategies},
\newblock \bibinfo{journal}{Applied Energy} \bibinfo{volume}{137}
  (\bibinfo{year}{2015}) \bibinfo{pages}{854--866}. \URLprefix
  \url{https://www.sciencedirect.com/science/article/pii/S030626191400751X}.
  \DOIprefix\doi{https://doi.org/10.1016/j.apenergy.2014.07.068}.
\bibitem[{Palomba et~al.(2020)Palomba, Borri, Charalampidis, Frazzica, Cabeza,
  and Karellas}]{PALOMBA2020190}
\bibinfo{author}{V.~Palomba}, \bibinfo{author}{E.~Borri},
  \bibinfo{author}{A.~Charalampidis}, \bibinfo{author}{A.~Frazzica},
  \bibinfo{author}{L.~F. Cabeza}, \bibinfo{author}{S.~Karellas},
\newblock \bibinfo{title}{{Implementation of a solar-biomass system for
  multi-family houses: Towards 100\% renewable energy utilization}},
\newblock \bibinfo{journal}{Renewable Energy} \bibinfo{volume}{166}
  (\bibinfo{year}{2020}) \bibinfo{pages}{190--209}.
  \DOIprefix\doi{https://doi.org/10.1016/j.renene.2020.11.126}.
\bibitem[{McKenna et~al.(2017)McKenna, Merkel, and Fichtner}]{McKenna.2017}
\bibinfo{author}{R.~McKenna}, \bibinfo{author}{E.~Merkel},
  \bibinfo{author}{W.~Fichtner},
\newblock \bibinfo{title}{Energy autonomy in residential buildings: A
  techno-economic model-based analysis of the scale effects},
\newblock \bibinfo{journal}{Applied Energy} \bibinfo{volume}{189}
  (\bibinfo{year}{2017}) \bibinfo{pages}{800--815}.
  \DOIprefix\doi{10.1016/j.apenergy.2016.03.062}.
\bibitem[{Merkel et~al.(2017)Merkel, Kunze, McKenna, and
  Fichtner}]{merkel2017modellgestutzte}
\bibinfo{author}{E.~Merkel}, \bibinfo{author}{R.~Kunze},
  \bibinfo{author}{R.~McKenna}, \bibinfo{author}{W.~Fichtner},
\newblock \bibinfo{title}{Modellgest{\"u}tzte bewertung des
  kraft-w{\"a}rme-kopplungsgesetzes 2016 anhand ausgew{\"a}hlter
  anwendungsf{\"a}lle in wohngeb{\"a}uden},
\newblock \bibinfo{journal}{Zeitschrift f{\"u}r Energiewirtschaft}
  \bibinfo{volume}{41} (\bibinfo{year}{2017}) \bibinfo{pages}{1--22}.
  \DOIprefix\doi{10.1007/s12398-017-0193-z}.
\bibitem[{Harder and Durmaz(2020)}]{Harder.2020}
\bibinfo{author}{S.~Harder}, \bibinfo{author}{A.~Durmaz},
\newblock \bibinfo{title}{Wohnungswirtschaft 2.0--transformation vom vermieter
  zum integrierten dezentralen versorger},
\newblock in: \bibinfo{booktitle}{Realisierung Utility 4.0 Band 2},
  \bibinfo{publisher}{Springer}, \bibinfo{year}{2020}, pp.
  \bibinfo{pages}{829--853}.
\bibitem[{Manns(2019)}]{Manns.2019}
\bibinfo{author}{P.~Manns},
\newblock \bibinfo{title}{A techno-environmental analysis on the effect of
  combining ev charging stations with pv tenant law systems and subsidies on
  the reduction of greenhouse gas emissions},
\newblock \bibinfo{journal}{PRE-PRINT}  (\bibinfo{year}{2019}).
\bibitem[{Scheller et~al.(2017)Scheller, Reichelt, Dienst, Johanning,
  Reichardt, and Bruckner}]{Scheller.2017}
\bibinfo{author}{F.~Scheller}, \bibinfo{author}{D.~G. Reichelt},
  \bibinfo{author}{S.~Dienst}, \bibinfo{author}{S.~Johanning},
  \bibinfo{author}{S.~Reichardt}, \bibinfo{author}{T.~Bruckner},
\newblock \bibinfo{title}{Effects of implementing decentralized business models
  at a neighborhood energy system level: A model based cross-sectoral
  analysis},
\newblock in: \bibinfo{booktitle}{2017 14th International Conference on the
  European Energy Market (EEM)}, \bibinfo{year}{2017}, pp.
  \bibinfo{pages}{1--6}. \DOIprefix\doi{10.1109/EEM.2017.7981910}.
\bibitem[{Seim et~al.(2017)Seim, Scheller, G{\"o}tz, Kondziella, and
  Bruckner}]{Seim.2017}
\bibinfo{author}{S.~Seim}, \bibinfo{author}{F.~Scheller},
  \bibinfo{author}{M.~G{\"o}tz}, \bibinfo{author}{H.~Kondziella},
  \bibinfo{author}{T.~Bruckner},
\newblock \bibinfo{title}{Assessment of pv-based business models in urban
  energy systems with respect to political and economic targets: A model-based
  scenario analysis},
\newblock \bibinfo{journal}{Internationale Energiewirtschaftstagung an der TU
  Wien}  (\bibinfo{year}{2017}).
\bibitem[{Knoop et~al.(2018)Knoop, Littwin, Kesting, and Ohrdes}]{Knoop.2018}
\bibinfo{author}{M.~Knoop}, \bibinfo{author}{M.~Littwin},
  \bibinfo{author}{M.~Kesting}, \bibinfo{author}{T.~Ohrdes},
  \bibinfo{title}{Modell zur {\"o}konomischen und {\"o}kologischen bewertung
  von geb{\"a}udeversorgungsverfahren im rahmen des mieterstromgesetzes -
  langfassung}, \bibinfo{year}{2018}, \bibinfo{editor}{{Photovoltaik-Symposium
  2018}} (Ed.).
\bibitem[{Frieden et~al.(2019)Frieden, Roberts, and Gubina}]{Frieden.2019}
\bibinfo{author}{D.~Frieden}, \bibinfo{author}{J.~Roberts},
  \bibinfo{author}{A.~F. Gubina},
\newblock \bibinfo{title}{Overview of emerging regulatory frameworks on
  collective self-consumption and energy communities in europe},
\newblock \bibinfo{journal}{2019 16th International Conference on the European
  Energy Market (EEM)}  (\bibinfo{year}{2019}).
  \DOIprefix\doi{https://doi.org/10.1109/EEM.2019.8916222}.
\bibitem[{Mohammadi et~al.(2017)Mohammadi, Noorollahi, Mohammadi-Ivatloo, and
  Yousefi}]{Mohammadi.2017}
\bibinfo{author}{M.~Mohammadi}, \bibinfo{author}{Y.~Noorollahi},
  \bibinfo{author}{B.~Mohammadi-Ivatloo}, \bibinfo{author}{H.~Yousefi},
\newblock \bibinfo{title}{Energy hub: From a model to a concept--a review},
\newblock \bibinfo{journal}{Renewable and Sustainable Energy Reviews}
  \bibinfo{volume}{80} (\bibinfo{year}{2017}) \bibinfo{pages}{1512--1527}.
  \DOIprefix\doi{10.1016/j.rser.2017.07.030}.
\bibitem[{Ghorab(2019)}]{Ghorab.2019}
\bibinfo{author}{M.~Ghorab},
\newblock \bibinfo{title}{Energy hubs optimization for smart energy network
  system to minimize economic and environmental impact at canadian community},
\newblock \bibinfo{journal}{Applied Thermal Engineering} \bibinfo{volume}{151}
  (\bibinfo{year}{2019}) \bibinfo{pages}{214--230}.
  \DOIprefix\doi{10.1016/j.applthermaleng.2019.01.107}.
\bibitem[{Orehounig et~al.(2015)Orehounig, Evins, and Dorer}]{Orehounig.2015}
\bibinfo{author}{K.~Orehounig}, \bibinfo{author}{R.~Evins},
  \bibinfo{author}{V.~Dorer},
\newblock \bibinfo{title}{Integration of decentralized energy systems in
  neighbourhoods using the energy hub approach},
\newblock \bibinfo{journal}{Applied Energy} \bibinfo{volume}{154}
  (\bibinfo{year}{2015}) \bibinfo{pages}{277--289}.
  \DOIprefix\doi{10.1016/j.apenergy.2015.04.114}.
\bibitem[{Falke et~al.(2016)Falke, Krengel, Meinerzhagen, and
  Schnettler}]{Falke.2016}
\bibinfo{author}{T.~Falke}, \bibinfo{author}{S.~Krengel},
  \bibinfo{author}{A.-K. Meinerzhagen}, \bibinfo{author}{A.~Schnettler},
\newblock \bibinfo{title}{Multi-objective optimization and simulation model for
  the design of distributed energy systems},
\newblock \bibinfo{journal}{Applied Energy} \bibinfo{volume}{184}
  (\bibinfo{year}{2016}) \bibinfo{pages}{1508--1516}. \URLprefix
  \url{http://www.sciencedirect.com/science/article/pii/S0306261916303646}.
  \DOIprefix\doi{10.1016/j.apenergy.2016.03.044}.
\bibitem[{Bati{\'c} et~al.(2016)Bati{\'c}, Toma{\v{s}}evi{\'c}, Beccuti,
  Demiray, and Vrane{\v{s}}}]{Batic.2016}
\bibinfo{author}{M.~Bati{\'c}}, \bibinfo{author}{N.~Toma{\v{s}}evi{\'c}},
  \bibinfo{author}{G.~Beccuti}, \bibinfo{author}{T.~Demiray},
  \bibinfo{author}{S.~Vrane{\v{s}}},
\newblock \bibinfo{title}{Combined energy hub optimisation and demand side
  management for buildings},
\newblock \bibinfo{journal}{Energy and Buildings} \bibinfo{volume}{127}
  (\bibinfo{year}{2016}) \bibinfo{pages}{229--241}.
  \DOIprefix\doi{10.1016/j.enbuild.2016.05.087}.
\bibitem[{Ma et~al.(2018)Ma, Xue, Liu, and Zhou}]{Ma.2018}
\bibinfo{author}{W.~Ma}, \bibinfo{author}{X.~Xue}, \bibinfo{author}{G.~Liu},
  \bibinfo{author}{R.~Zhou},
\newblock \bibinfo{title}{Techno-economic evaluation of a community-based
  hybrid renewable energy system considering site-specific nature},
\newblock \bibinfo{journal}{Energy Conversion and Management}
  \bibinfo{volume}{171} (\bibinfo{year}{2018}) \bibinfo{pages}{1737--1748}.
  \DOIprefix\doi{10.1016/j.enconman.2018.06.109}.
\bibitem[{Jing et~al.(2020)Jing, Xie, Wang, and Chen}]{Jing.2020}
\bibinfo{author}{R.~Jing}, \bibinfo{author}{M.~N. Xie}, \bibinfo{author}{F.~X.
  Wang}, \bibinfo{author}{L.~X. Chen},
\newblock \bibinfo{title}{Fair p2p energy trading between residential and
  commercial multi-energy systems enabling integrated demand-side management},
\newblock \bibinfo{journal}{Applied Energy} \bibinfo{volume}{262}
  (\bibinfo{year}{2020}) \bibinfo{pages}{114551}.
  \DOIprefix\doi{10.1016/j.apenergy.2020.114551}.
\bibitem[{Eshraghi et~al.(2019)Eshraghi, Salehi, Heibati, and
  Lari}]{Eshraghi.2019}
\bibinfo{author}{A.~Eshraghi}, \bibinfo{author}{G.~Salehi},
  \bibinfo{author}{S.~Heibati}, \bibinfo{author}{K.~Lari},
\newblock \bibinfo{title}{Developing operation of combined cooling, heat, and
  power system based on energy hub in a micro-energy grid: The application of
  energy storages},
\newblock \bibinfo{journal}{Energy {\&} Environment} \bibinfo{volume}{30}
  (\bibinfo{year}{2019}) \bibinfo{pages}{1356--1379}.
  \DOIprefix\doi{10.1177/0958305X19846577}.
\bibitem[{Scheller et~al.(2020)Scheller, Burkhardt, Schwarzeit, McKenna, and
  Bruckner}]{Scheller.2020}
\bibinfo{author}{F.~Scheller}, \bibinfo{author}{R.~Burkhardt},
  \bibinfo{author}{R.~Schwarzeit}, \bibinfo{author}{R.~McKenna},
  \bibinfo{author}{T.~Bruckner},
\newblock \bibinfo{title}{Competition between simultaneous demand-side
  flexibility options: The case of community electricity storage systems},
\newblock \bibinfo{journal}{Applied Energy} \bibinfo{volume}{269}
  (\bibinfo{year}{2020}) \bibinfo{pages}{114969}.
  \DOIprefix\doi{10.1016/j.apenergy.2020.114969}.
\bibitem[{Scheller et~al.(2018{\natexlab{a}})Scheller, Johanning, Reichardt,
  Reichelt, and Bruckner}]{Scheller.2018b}
\bibinfo{author}{F.~Scheller}, \bibinfo{author}{S.~Johanning},
  \bibinfo{author}{S.~Reichardt}, \bibinfo{author}{D.~G. Reichelt},
  \bibinfo{author}{T.~Bruckner},
\newblock \bibinfo{title}{Competition effects of simultaneous application of
  flexibility options within an energy community},
\newblock in: \bibinfo{booktitle}{2018 15th International Conference on the
  European Energy Market (EEM)}, \bibinfo{year}{2018}{\natexlab{a}}, pp.
  \bibinfo{pages}{1--5}. \DOIprefix\doi{10.1109/EEM.2018.8470007}.
\bibitem[{Scheller et~al.(2018{\natexlab{b}})Scheller, Burgenmeister,
  Kondziella, K{\"u}hne, Reichelt, and Bruckner}]{scheller2018towards}
\bibinfo{author}{F.~Scheller}, \bibinfo{author}{B.~Burgenmeister},
  \bibinfo{author}{H.~Kondziella}, \bibinfo{author}{S.~K{\"u}hne},
  \bibinfo{author}{D.~G. Reichelt}, \bibinfo{author}{T.~Bruckner},
\newblock \bibinfo{title}{Towards integrated multi-modal municipal energy
  systems: An actor-oriented optimization approach},
\newblock \bibinfo{journal}{Applied Energy} \bibinfo{volume}{228}
  (\bibinfo{year}{2018}{\natexlab{b}}) \bibinfo{pages}{2009--2023}.
  \DOIprefix\doi{10.1016/j.apenergy.2018.07.027}.
\bibitem[{Scheller et~al.(2018{\natexlab{c}})Scheller, Johanning, Seim,
  Schuchardt, Krone, Haberland, and Bruckner}]{Scheller.2018}
\bibinfo{author}{F.~Scheller}, \bibinfo{author}{S.~Johanning},
  \bibinfo{author}{S.~Seim}, \bibinfo{author}{K.~Schuchardt},
  \bibinfo{author}{J.~Krone}, \bibinfo{author}{R.~Haberland},
  \bibinfo{author}{T.~Bruckner},
\newblock \bibinfo{title}{Legal framework of decentralized energy business
  models in germany: Challenges and opportunities for municipal utilities},
\newblock \bibinfo{journal}{Zeitschrift f{\"u}r Energiewirtschaft}
  \bibinfo{volume}{42} (\bibinfo{year}{2018}{\natexlab{c}})
  \bibinfo{pages}{207--223}. \DOIprefix\doi{10.1007/s12398-018-0227-1}.
\bibitem[{Braeuer et~al.(2019)Braeuer, Kleinebrahm, and Naber}]{Braeuer.2019}
\bibinfo{author}{F.~Braeuer}, \bibinfo{author}{M.~Kleinebrahm},
  \bibinfo{author}{E.~Naber},
\newblock \bibinfo{title}{Effects of the tenants electricity law on energy
  system layout and landlord-tenant relationship in a multi-family building in
  germany},
\newblock \bibinfo{journal}{IOP Conference Series: Earth and Environmental
  Science} \bibinfo{volume}{323} (\bibinfo{year}{2019})
  \bibinfo{pages}{012168}. \DOIprefix\doi{10.1088/1755-1315/323/1/012168}.
\bibitem[{{Bundesministerium f{\"u}r Wirtschaft und Energie}(2017)}]{EEG.2017a}
\bibinfo{author}{{Bundesministerium f{\"u}r Wirtschaft und Energie}},
  \bibinfo{title}{Gesetz zur f{\"o}rderung von mieterstrom und zur {\"a}nderung
  weiterer vorschriften des erneuerbare-energien-gesetzes},
  \bibinfo{year}{17.07.2017}. \URLprefix
  \url{https://www.bundesnetzagentur.de/SharedDocs/Downloads/DE/Sachgebiete/Energie/Verbraucher/Vertragsarten/Mieterstrom_BGBl.pdf?__blob=publicationFile&v=2}.
\bibitem[{{Bundesministerium f{\"u}r Wirtschaft und Energie}(2021)}]{EEG.2021}
\bibinfo{author}{{Bundesministerium f{\"u}r Wirtschaft und Energie}},
  \bibinfo{title}{Gesetz f{\"u}r den ausbau erneuerbarer energien
  (erneuerbare-energien-gesetz - eeg 2021): Eeg 2021}, \bibinfo{year}{2021}.
  \URLprefix \url{https://www.gesetze-im-internet.de/eeg_2014/EEG_2021.pdf}.
\bibitem[{Herz and Henning(2018)}]{Herz.2018}
\bibinfo{author}{S.~Herz}, \bibinfo{author}{B.~Henning},
  \bibinfo{title}{Rechtsgutachten: ``kleiner mieterstrom'' und
  gemeinschaftliche eigenversorgung}, \bibinfo{year}{2018}. \URLprefix
  \url{https://www.verbraucherzentrale.nrw/sites/default/files/2019-01/Rechtsgutachten\%20Gemeinschaftliche\%20Eigenversorgung.pdf}.
\bibitem[{Behr and Gro{\ss}klos(2017)}]{Behr.2017}
\bibinfo{editor}{I.~Behr}, \bibinfo{editor}{M.~Gro{\ss}klos} (Eds.),
  \bibinfo{title}{Praxishandbuch Mieterstrom : Fakten, Argumente und
  Strategien}, \bibinfo{publisher}{{Springer Fachmedien Wiesbaden}},
  \bibinfo{address}{Wiesbaden}, \bibinfo{year}{2017}.
\bibitem[{{Bundesministerium f{\"u}r Wirtschaft und Energie}(2015)}]{KWKG.2018}
\bibinfo{author}{{Bundesministerium f{\"u}r Wirtschaft und Energie}},
  \bibinfo{title}{Gesetz f{\"u}r die erhaltung, die modernisierung und den
  ausbau der kraft-w{\"a}rme-kopplung (kraft-w{\"a}rme-kopplungsgesetz): Kwkg},
  \bibinfo{year}{2015}.
\bibitem[{{Bundesministerium f{\"u}r Wirtschaft und Energie}(2020)}]{KWKG.2020}
\bibinfo{author}{{Bundesministerium f{\"u}r Wirtschaft und Energie}},
  \bibinfo{title}{Gesetz f{\"u}r die erhaltung, die modernisierung und den
  ausbau der kraft-w{\"a}rme-kopplung (kraft-w{\"a}rme-kopplungsgesetz): Kwkg},
  \bibinfo{year}{2020}.
\bibitem[{Briem et~al.(2020)Briem, Beckers, Bunkus, Fabris, Hoffmann,
  Herbsttritt, Hofmeier, Krack, Nowack, Rother, Schuberth, Steinbrenner,
  Sternkopf, Unnterstall, and Vollmer}]{UBA.2020}
\bibinfo{author}{S.~Briem}, \bibinfo{author}{R.~Beckers},
  \bibinfo{author}{R.~Bunkus}, \bibinfo{author}{C.~Fabris},
  \bibinfo{author}{F.~Hoffmann}, \bibinfo{author}{C.~Herbsttritt},
  \bibinfo{author}{K.~Hofmeier}, \bibinfo{author}{J.~Krack},
  \bibinfo{author}{A.~Nowack}, \bibinfo{author}{S.~Rother},
  \bibinfo{author}{J.~Schuberth}, \bibinfo{author}{J.~Steinbrenner},
  \bibinfo{author}{R.~Sternkopf}, \bibinfo{author}{H.~Unnterstall},
  \bibinfo{author}{C.~Vollmer}, \bibinfo{title}{Status quo der
  kraft-w{\"a}rme-kopplung in deutschland}, \bibinfo{year}{2020},
  \bibinfo{editor}{Umweltbundesamt} (Ed.). \URLprefix
  \url{www.umweltbundesamt.de/publikationen}.
\bibitem[{Kleinebrahm et~al.(2018)Kleinebrahm, Weinand, Ardone, and
  McKenna}]{Kleinebrahm.2018}
\bibinfo{author}{M.~Kleinebrahm}, \bibinfo{author}{J.~Weinand},
  \bibinfo{author}{A.~Ardone}, \bibinfo{author}{R.~McKenna},
  \bibinfo{title}{Optimal renewable energy based supply systems for
  self-sufficient residential buildings}, \bibinfo{year}{2018}.
  \DOIprefix\doi{10.5445/IR/1000085753}.
\bibitem[{{Bundesministerium f{\"u}r Wirtschaft und Energie}(2020)}]{BMWi.2020}
\bibinfo{author}{{Bundesministerium f{\"u}r Wirtschaft und Energie}},
  \bibinfo{title}{Long-term renovation strategy of the federal government},
  \bibinfo{year}{2020}, \bibinfo{editor}{{Bundesministerium f{\"u}r Wirtschaft
  und Energie}} (Ed.). \URLprefix
  \url{https://ec.europa.eu/energy/sites/default/files/documents/de_2020_ltrs_official_en_translation.pdf}.
\bibitem[{Ziesing et~al.(2019)Ziesing, Maa{\ss}en, and Nicker}]{AGEB.2019}
\bibinfo{author}{H.-J. Ziesing}, \bibinfo{author}{U.~Maa{\ss}en},
  \bibinfo{author}{M.~Nicker}, \bibinfo{title}{Energie in zahlen: Arbeit und
  leistungen der ag energiebilanzen}, \bibinfo{year}{2019},
  \bibinfo{editor}{{Arbeitsgemeinschaft Energiebilanzen e.V.}} (Ed.).
\bibitem[{Bundesnetzagentur and
  Bundeskartellamt(2020)}]{Bundesnetzagentur.2020b}
\bibinfo{author}{Bundesnetzagentur}, \bibinfo{author}{Bundeskartellamt},
  \bibinfo{title}{Monitoringbericht energie 2020}, \bibinfo{year}{2020},
  \bibinfo{editor}{Bundesnetzagentur} (Ed.). \URLprefix
  \url{https://www.bundesnetzagentur.de/SharedDocs/Mediathek/Berichte/2020/Monitoringbericht_Energie2020.pdf?__blob=publicationFile&v=5}.
\bibitem[{{Fraktionen der CDU/CSU und SPD}(2019)}]{BEHG.2019}
\bibinfo{author}{{Fraktionen der CDU/CSU und SPD}}, \bibinfo{title}{Gesetz
  {\"u}ber einen nationalen zertifikatehandel f{\"u}r brennstoffemissionen:
  Brennstoffemissionshandelsgesetz - behg}, \bibinfo{year}{2019}. \URLprefix
  \url{http://www.gesetze-im-internet.de/behg/BEHG.pdf}.
\bibitem[{Fischer et~al.(2015)Fischer, H{\"a}rtl, and
  Wille-Haussmann}]{Fischer.2015}
\bibinfo{author}{D.~Fischer}, \bibinfo{author}{A.~H{\"a}rtl},
  \bibinfo{author}{B.~Wille-Haussmann},
\newblock \bibinfo{title}{Model for electric load profiles with high time
  resolution for german households},
\newblock \bibinfo{journal}{Energy and Buildings} \bibinfo{volume}{92}
  (\bibinfo{year}{2015}) \bibinfo{pages}{170--179}.
  \DOIprefix\doi{10.1016/j.enbuild.2015.01.058}.
\bibitem[{Kaschub et~al.(2016)Kaschub, Jochem, and Fichtner}]{Kaschub.2016}
\bibinfo{author}{T.~Kaschub}, \bibinfo{author}{P.~Jochem},
  \bibinfo{author}{W.~Fichtner},
\newblock \bibinfo{title}{Solar energy storage in german households:
  Profitability, load changes and flexibility},
\newblock \bibinfo{journal}{Energy Policy} \bibinfo{volume}{98}
  (\bibinfo{year}{2016}) \bibinfo{pages}{520--532}.
  \DOIprefix\doi{10.1016/j.enpol.2016.09.017}.
\bibitem[{Loga et~al.(2015)Loga, Stein, Diefenbach, and Born}]{IWU.2015}
\bibinfo{author}{T.~Loga}, \bibinfo{author}{B.~Stein},
  \bibinfo{author}{N.~Diefenbach}, \bibinfo{author}{R.~Born},
  \bibinfo{title}{Detsche wohngeb{\"a}udetypologie: Beispielhafte ma{\ss}nahmen
  zur verbesserung der energieeiffiezienz von typischen wohngeb{\"a}uden},
  \bibinfo{year}{2015}, \bibinfo{editor}{IWU} (Ed.).
\bibitem[{{Deutscher Bundestag}(2021)}]{Bundestag.2021}
\bibinfo{author}{{Deutscher Bundestag}}, \bibinfo{title}{Plenarprotokoll
  19/208: Stenografischer bericht 208. sitzung}, \bibinfo{year}{10.02.2021},
  \bibinfo{editor}{{Deutscher Bundestag}} (Ed.). \URLprefix
  \url{https://dserver.bundestag.de/btp/19/19208.pdf}.
\bibitem[{McKenna(2018)}]{McKenna.2018}
\bibinfo{author}{R.~McKenna},
\newblock \bibinfo{title}{The double-edged sword of decentralized energy
  autonomy},
\newblock \bibinfo{journal}{Energy Policy} \bibinfo{volume}{113}
  (\bibinfo{year}{2018}) \bibinfo{pages}{747--750}.
  \DOIprefix\doi{10.1016/j.enpol.2017.11.033}.
\bibitem[{McKenna et~al.(2019)McKenna, Fehrenbach, and Merkel}]{McKenna.2019}
\bibinfo{author}{R.~McKenna}, \bibinfo{author}{D.~Fehrenbach},
  \bibinfo{author}{E.~Merkel},
\newblock \bibinfo{title}{The role of seasonal thermal energy storage in
  increasing renewable heating shares: A techno-economic analysis for a typical
  residential district},
\newblock \bibinfo{journal}{Energy and Buildings} \bibinfo{volume}{187}
  (\bibinfo{year}{2019}) \bibinfo{pages}{38--49}.
  \DOIprefix\doi{10.1016/j.enbuild.2019.01.044}.
\bibitem[{McKenna et~al.(2016)McKenna, Hofmann, Merkel, Fichtner, and
  Strachan}]{McKenna.2016}
\bibinfo{author}{R.~McKenna}, \bibinfo{author}{L.~Hofmann},
  \bibinfo{author}{E.~Merkel}, \bibinfo{author}{W.~Fichtner},
  \bibinfo{author}{N.~Strachan},
\newblock \bibinfo{title}{Analysing socioeconomic diversity and scaling effects
  on residential electricity load profiles in the context of low carbon
  technology uptake},
\newblock \bibinfo{journal}{Energy Policy} \bibinfo{volume}{97}
  (\bibinfo{year}{2016}) \bibinfo{pages}{13--26}.
  \DOIprefix\doi{10.1016/j.enpol.2016.06.042}.

\end{thebibliography}


\begin{thebibliography}{3}
\expandafter\ifx\csname natexlab\endcsname\relax\def\natexlab#1{#1}\fi
\providecommand{\url}[1]{\texttt{#1}}
\providecommand{\href}[2]{#2}
\providecommand{\path}[1]{#1}
\providecommand{\DOIprefix}{doi:}
\providecommand{\ArXivprefix}{arXiv:}
\providecommand{\URLprefix}{URL: }
\providecommand{\Pubmedprefix}{pmid:}
\providecommand{\doi}[1]{\href{http://dx.doi.org/#1}{\path{#1}}}
\providecommand{\Pubmed}[1]{\href{pmid:#1}{\path{#1}}}
\providecommand{\bibinfo}[2]{#2}
\ifx\xfnm\relax \def\xfnm[#1]{\unskip,\space#1}\fi
\bibitem[{Ziesing et~al.(2019)Ziesing, Maa{\ss}en, and Nicker}]{AGEB.2019}
\bibinfo{author}{H.-J. Ziesing}, \bibinfo{author}{U.~Maa{\ss}en},
  \bibinfo{author}{M.~Nicker}, \bibinfo{title}{Energie in zahlen: Arbeit und
  leistungen der ag energiebilanzen}, \bibinfo{year}{2019},
  \bibinfo{editor}{{Arbeitsgemeinschaft Energiebilanzen e.V.}} (Ed.).
\bibitem[{{European Parliament and the Council}(2012)}]{EU.2012}
\bibinfo{author}{{European Parliament and the Council}},
  \bibinfo{title}{Directive 2012/27/eu on on energy efficiency, amending
  directives 2009/125/ec and 2010/30/eu and repealing directives 2004/8/ec and
  2006/32/ec}, \bibinfo{year}{2012}. \URLprefix
  \url{https://eur-lex.europa.eu/legal-content/EN/TXT/PDF/?uri=CELEX:02012L0027-20210101&qid=1616156929004&from=en}.
\bibitem[{{The European Comission}(2015)}]{EU.2015}
\bibinfo{author}{{The European Comission}}, \bibinfo{title}{Commission
  delegated regulation (eu) 2015/ 2402 - of 12 october 2015 - reviewing
  harmonised efficiency reference values for separate production of electricity
  and heat in application of directive 2012/ 27/ eu of the european parliament
  and of the council and repealing commission implementing decision 2011/ 877/
  eu}, \bibinfo{year}{2015}. \URLprefix
  \url{https://eur-lex.europa.eu/legal-content/EN/TXT/PDF/?uri=CELEX:32015R2402&from=de}.

\end{thebibliography}




\clearpage
\pagenumbering{arabic}
    \centerline{\textbf{Supplementary Information}}
    \centerline{for}
    
    \begin{center}
    \textbf{Optimal system design for energy communities in multi-family buildings: the case of the German Tenant Electricity Law}%
    \newline
    \newline
    Fritz Braeuer, \textit{Institute for Industrial Production (IIP), Karlsruhe Institute of Technology (KIT), Karlsruhe, Germany}
    
    Max Kleinebrahm, \textit{Institute for Industrial Production (IIP), Karlsruhe Institute of Technology (KIT), Karlsruhe, Germany}
    
    Elias Naber, \textit{Institute for Industrial Production (IIP), Karlsruhe Institute of Technology (KIT), Karlsruhe, Germany}
    
    Fabian Scheller, \textit{Energy Systems Analysis, Division of Sustainability, Technical University of Denmark (DTU), Kgs. Lyngby, Denmark}
    
    Russell McKenna, \textit{Chair of Energy Transition, School of Engineering, University of Aberdeen, Aberdeen, United Kingdom}

\end{center}
\appendix
\renewcommand\appendixname{SI}

\section{Emissions for electricity fed into the grid}
\label{App_emissionsintogrid}

Regarding the allocation of fuel and  emissions respectively from co-generation processes in the CHP-unit the alternative generation method is used \ref{eq_chp_allocation}. This method is utilized for the national and official energy accounting in Germany \citeSI{AGEB.2019}. Furthermore, it is in line with the methodology for determining the efficiency of the co-generation process in \citeSI{EU.2012} and \citeSI{EU.2015}. According to the national Working Group on Energy Balances (AGEB) \citeSI{AGEB.2019} parameters for the alternative efficiencies are $\eta_{th,alt}=0.80$ and $\eta_{el,alt}=0.40$\footnote{Based on \citeSI{EU.2015} the parameters would be $\eta_{th,alt}=0.92$ and $\eta_{el,alt}=0.455$ which would alter the specific emissions coefficients by +/- 1g each}. 

\begin{equation}
\label{eq_chp_allocation}
\begin{split}
EF_{chp,el}&=(1-PES)*\frac{\eta_{chp,el}}{\eta_{el,alt}} \cdot \frac{EF_{gas}}{\eta_{chp,el}}  \;\;   with  \;\; PES=1-\frac{1}{\frac{\eta_{chp,th}}{\eta_{th,alt}}+\frac{\eta_{chp,el}}{\eta_{el,alt}}}\\
\end{split}
\end{equation}

\section{In-house cash-flow and energy flow}
\label{App_InhouseCashFlow}

Equations \ref{eq_returnCHPfeedin}, \ref{eq_returnCHPtenant} and \ref{eq_returnCHPHP} further explain the calculation of earnings from CHP operation. In Equation \ref{eq_returnCHPfeedin} the CHP unit generates heat and feeds the electricity into the grid, thus only payment for heat and through the feed-in tarif can be charged. Equation \ref{eq_returnCHPtenant} displays the operational mode where the electricity of the CHP unit is sold to the tenants and the SCP and the tenant electricity price is charged. Finally, Equation \ref{eq_returnCHPHP} represents the case, where the electricity is converted in the HP to heat. This increases the amount of heat sold to tenants and charges only the SCP for electricity self-consumption.

\begin{align}
\label{eq_returnCHPfeedin}
\begin{split}
    R_{chp,feedin} &= \frac{1}{\sigma_{chp}} \cdot \frac{c_{gas}}{\eta_{boiler}} + c_{chp,feedin} - \frac{c_{gas}}{\eta_{el,chp}} \\
      &= \Big[ \frac{1}{0.6} \cdot \frac{6.33}{0.85} + 16.00 - \frac{6.33}{0.35} \Big] \; \mathit{ct} \slash \mathit{kWh_{el}} \\ 
      &= 10.33 \; \mathit{ct} \slash \mathit{kWh_{el}}
\end{split}
\end{align}

\begin{align}
\label{eq_returnCHPtenant}
\begin{split}
    R_{chp,te} &= \frac{1}{\sigma_{chp}} \cdot \frac{c_{gas}}{\eta_{boiler}} + (c_{el,te} + c_{chp,te} - c_{fees}) - \frac{c_{gas}}{\eta_{el,chp}} \\
      &= \Big[ \frac{1}{0.6} \cdot \frac{6.33}{0.85} + (31.03 + 8.00 - 12.08) - \frac{6.33}{0.35} \Big] \; \mathit{ct} \slash \mathit{kWh_{el}} \\ 
      &= 21.28 \; \mathit{ct} \slash \mathit{kWh_{el}}
\end{split}
\end{align}

\begin{align}
\label{eq_returnCHPHP}
\begin{split}
  R_{chp,te} &= (\frac{1}{\sigma_{chp}} + \mathit{COP_{hp}} ) \cdot \frac{c_{gas}}{\eta_{boiler}} + (c_{chp,te} - 0.4 * c_{chp,self}) - \frac{c_{gas}}{\eta_{el,chp}} \\
      &= \Big[ (\frac{1}{0.6} + 3.5) \cdot \frac{6.33}{0.85} + (8.00 - 0.4 \cdot 6.5) - \frac{6.33}{0.35} \Big] \; \mathit{ct} \slash \mathit{kWh_{el}} \\ 
      &= 25.79 \; \mathit{ct} \slash \mathit{kWh_{el}}
\end{split}
\end{align}

\paragraph{Energy flow illustration}

\begin{figure}[h]
\centering
\includegraphics[width=\textwidth]{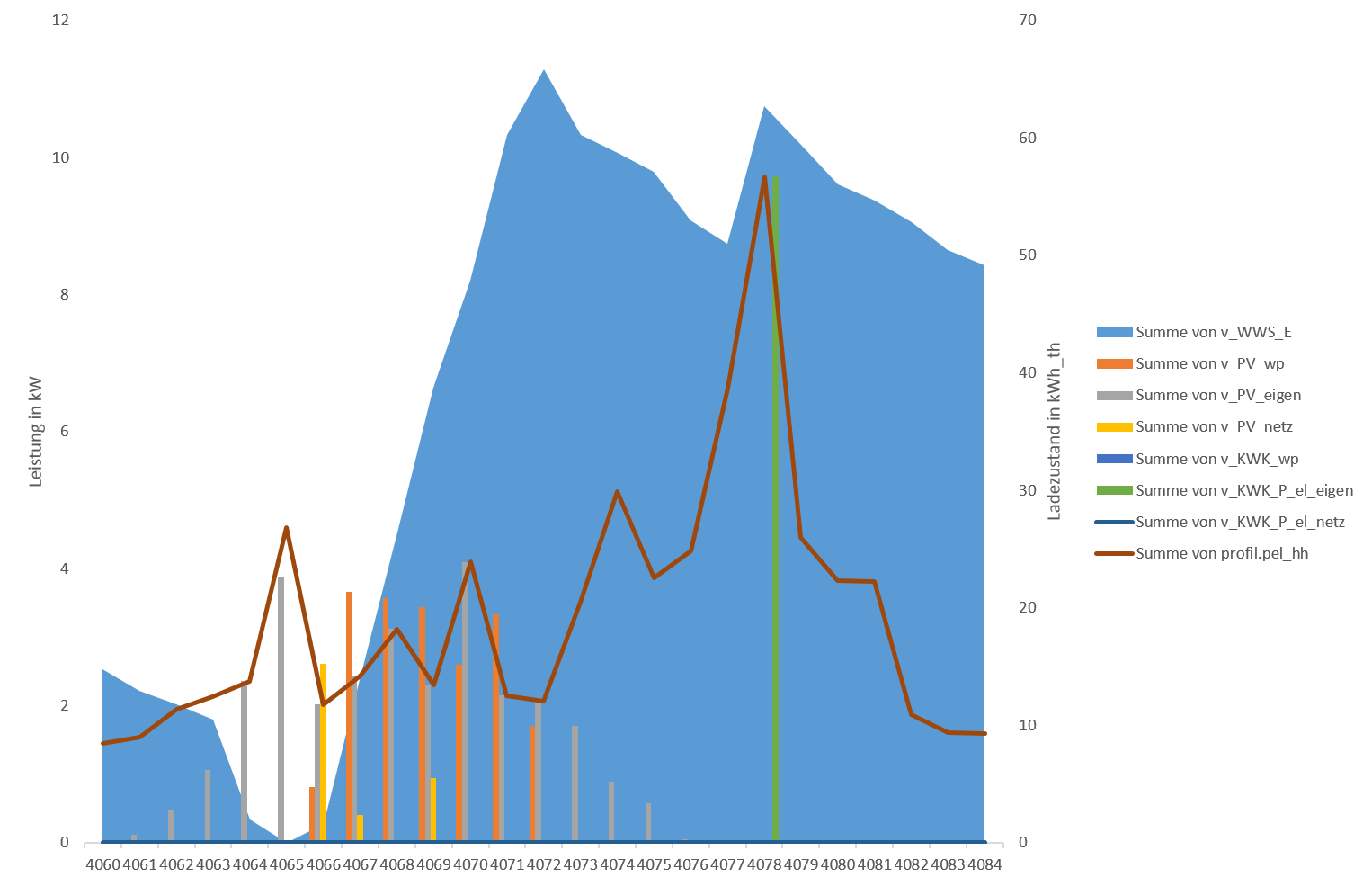}
\caption{Exemplary energy flow for day 1 in summer for Building 1, system \textit{all} and a CHP unit of \kWel{20}.}
\label{fig.exempleflow1}
\end{figure}

\begin{figure}[h]
\centering
\includegraphics[width=\textwidth]{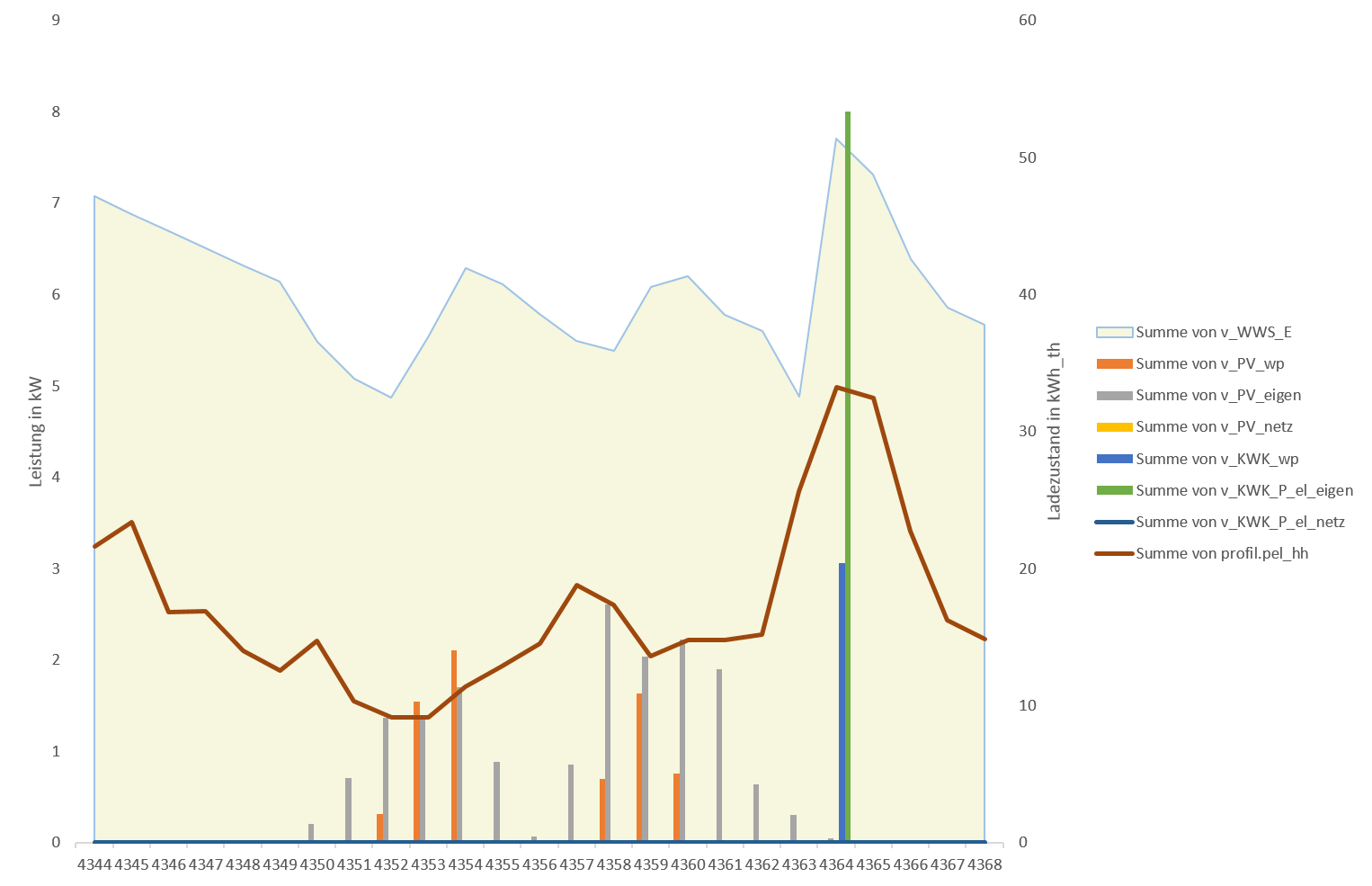}
\caption{Exemplary energy flow for day 1 in summer for Building 1, system \textit{all} and a CHP unit of \kWel{20}.}
\label{fig.exempleflow2}
\end{figure}

Figure \ref{fig.exempleflow1} and Figure \ref{fig.exempleflow2} show the energy flows for two days in summer. It shows the system \textit{all} for building 1 where the CHP unit is \kWel{20} large. The model has chosen not to install a boiler, which results in the CHP and the HP as the only heat source in the building. Figure \ref{fig.exempleflow1} illustrates that most of the PV electricity is self-consumed within the building. In this case the minimum load of the CHP is \kWel{8}. Thus, the electricity load during the day is too low to be covered by the CHP. One part of the PV electricity is directed to the HP to generate heat and fill the heat storage. The potential of heat generation of the HP is limited by the size of the heat storage and the operation of the CHP unit in the evening. In the evening there is a peak demand for electricity, which allows the CHP to operate. In order to do so, the heat as a bi-product of the electricity generation needs to be either stored or directly consumed; storage capacity and heat demand are limiting factors. This illustration helps to grasp the complexity of identifying the optimal dispatch. It also shows the effect of perfect foresight on the dispatch decision.

Figure \ref{fig.exempleflow2} shows another exemplary summer day. Here, the storage level is fairly high, which is a result of a dispatch decision in the following day. The figure illustrates the operational challenges of the CHP unit. It only operates during one hour in the evening to cover the peak demand. The electricity demand of the households during this hour is still below the minimum load of the CHP. Thus, the model chooses to generate additional electricity that is directed to the HP in order to surpass the minimum load restriction. Notably, the heat storage with around \kWh{66} is not fully utilized. This indicates that the dispatch of CHP and HP is additionally constrained by the heat demand of the building, which is relatively low during the summer season.

\newpage

\section{Cascading mode for the CHP unit}
From the results of the main analysis, we understand that the minimum load criteria of the CHP operation ($P_{chp,min}$) influences the system's dimensions as well as the dispatch decisions. To calculate the optimal dispatch of the CHP unit presents substantial complexities. To reduce the complexity of the problem, it is possible to aggregate the time steps or introduce other simplifications or assumptions. One simplification, which offers additional economic implications, is the assumption of a cascading mode in the CHP installation. The cascading mode describes the possibility to install multiple smaller CHP-units instead of one large unit. Thus, the system can operate continuously between a relatively small minimum load and its maximum load.  For example, considering a minimum load of \kWel{4}, one would install a first CHP unit with \kWel{10}. In a cascading mode to guarantee a continuous operation, the second CHP unit would need to provide a minimum load of \kWel{6} with a capacity of \kWel{15}. The next step is to invest in a \kWel{22.5} CHP unit, thus achieving a full capacity of \kWel{47.5} with a minimum load of \kWel{4}. 

\begin{equation}
    cap_{chp,1} = 4\,\mathit{kW_{el}}  + P_{chp,min,2} = 10\,\mathit{kW_{el}}
\end{equation}

\begin{align}
\begin{split}
    cap_{chp,1} + cap_{chp,2} &= cap_{chp,min,1} + P_{chp,2} + P_{chp,min,3} \\
    25\,\mathit{kW_{el}} &= 10\,\mathit{kW_{el}}  + 6\,\mathit{kW_{el}} + P_{chp,min,3} 
\end{split}
\end{align}

To implement the cascading mode into the model, we doubled the fix cost for the CHP system or tripled it for a cascading mode of two units or three units respectively. This price variation can also be interpreted as investment in other CHP technologies that are more expensive. Nonetheless, for the analysis of the cascading mode, the minimum load was set to \kWel{4} and the CHP capacity is implemented as a continuous variable of the optimization model. All the other CHP parameters are not changed. Table \ref{tbl_results_comp_2,0} and Table \ref{tbl_results_comp_3,0} present the results of the cascading mode analysis where the fixed price for a CHP unit is either doubled or tripled respectively. The results reveal that the price variation does not have an impact on the system design but on the NPV as the CHP investment increases. Compared to the main results in Table \ref{tbl_results_comp}, the model chooses to expand the CHP unit where possible. The CHP unit mostly operates within the subsidized 30,000 full load hours. As the CHP capacity is enlarged, the PV dimensions are smaller, \CO{} abatement is and the \textit{SCR} reduces as more electricity is fed into the grid.

The results are of economic relevance for a real life application. There exist technologies like natural gas driven fuel cells, that offer a lower minimum load than conventional CHPs. Nonetheless, the investment is much higher. The results in Table \ref{tbl_results_comp_2,0} and \ref{tbl_results_comp_3,0} indicate that the investment in a CHP unit with double or triple the fixed costs still yields a positive NPV and is favorable compared to a system without a CHP. As the optimal dispatch for a large CHP unit as seen in the main results is rather complex, it might be economically advantageous to pay the higher investment for a system with a low minimum load. This allows for a more flexible operation and avoids uncertainty risks.
\begin{table}[h] 
\centering 
\begin{tabular}{|lllllllll|}
\hline 
KPI & Unit &REF & CHP    & PV\_ & CHP\_ & COMBI & COMBI\_ & COMBI\_ \\ 
    &      &    &        & CHP   & HP     &       & EV       & EVopt    \\ 
\hline 
$NPV$ & \kEUR{} & -13.4 & 37.3 & 39.3 & 41.9 & 45.0 & 55.0 & 59.4 \\
$\Delta NPV$ & \kEUR{} & - & 50.7 & 52.7 & 55.3 & 58.4 & 68.5 & 72.8 \\
\hline 
\hline 
$D_{el,te}$ & MWh\slash a & 29.8 & 29.8 & 29.8 & 29.8 & 29.8 & 36.8 & 36.9 \\
$D_{el,hp}$ & MWh\slash a & - & - & - & 17.2 & 18.8 & 17.6 & 18.6 \\
$D_{el,tot}$ & MWh\slash a & 29.8 & 29.8 & 29.8 & 47.0 & 48.7 & 54.5 & 55.4 \\
\hline 
$cap_{PV}$ & kW & - & - & 5.8 & - & 8.1 & 12.0 & 12.7 \\
$cap_{CHP}$ & kWel & - & 50.0 & 50.0 & 32.8 & 30.8 & 32.3 & 31.1 \\
$cap_{HP}$ & kWth & - & - & - & 11.0 & 11.7 & 11.3 & 11.7 \\
$cap_{WWT}$ & kWh & - & 40.0 & 40.6 & 47.4 & 50.8 & 50.1 & 41.4 \\
$cap_{boil}$ & kWh & 60.3 & - & - & - & - & - & - \\
\hline 
$P_{chp,max}$ & kW & - & 40.1 & 40.1 & 32.8 & 30.8 & 32.3 & 31.1 \\
$h_{chp,full}$ & hours & - & 30184 & 30189 & 30000 & 30000 & 30000 & 30000 \\
\hline 
$SCR_{el}$ & \% & - & 32.6 & 33.2 & 85.0 & 85.1 & 85.4 & 91.2 \\
$DSS_{el}$ & \% & - & 82.5 & 89.9 & 88.9 & 93.7 & 92.8 & 95.5 \\
$DA_{el}$ & \% & - & 253.2 & 270.8 & 104.6 & 110.1 & 108.7 & 104.8 \\
$GII$ & \% & - & 19.5 & 19.6 & 10.6 & 11.2 & 10.9 & 9.9 \\
$GII_{norm}$ & \% & - & 125.5 & 126.5 & 57.4 & 60.3 & 80.0 & 94.0 \\
\hline 
$CO_{2,ref}$ & t & 659.3 & 659.3 & 659.3 & 659.3 & 659.3 & 688.9 & 688.9 \\
$CO_{2,opt}$ & t & 659.3 & 888.6 & 879.5 & 586.8 & 544.0 & 572.1 & 545.6 \\
$\Delta CO_{2}$ & t & - & -229.3 & -220.1 & 72.6 & 115.3 & 116.7 & 143.3 \\
\hline 
$CO_{2,export}$ & t & - & 318.7 & 334.1 & 46.3 & 44.8 & 43.0 & 32.0 \\
$\Delta CO_{2,export}$ & t & - & 105.4 & 108.0 & 15.3 & 11.4 & 6.8 & 10.6 \\
\hline 
$CF_{subs}$ & \kEUR{} & - & 136.9 & 142.6 & 61.5 & 61.9 & 66.1 & 61.6 \\
$cac_{subs}$ & \eurotCO{} & - & - & - & 847.7 & 536.8 & 566.5 & 430.1 \\
\hline 
\end{tabular}
\caption{Results for component wise analysis sorted by $\Delta NPV$, 
                cascading mode with double the fix price for CHP representing a system with two CHP-units} 
\label{tbl_results_comp_2,0}
\end{table}

\begin{table}[h] 
\centering 
\begin{tabular}{|lllllllll|}
\hline 
KPI & Unit &REF & CHP    & PV\_ & CHP\_ & COMBI & COMBI\_ & COMBI\_ \\ 
    &      &    &        & CHP   & HP     &       & EV       & EVopt    \\ 
\hline 
$NPV$ & \kEUR{} & -13.4 & 22.3 & 24.4 & 26.9 & 30.3 & 40.0 & 44.3 \\
$\Delta NPV$ & \kEUR{} & - & 35.7 & 37.8 & 40.3 & 43.7 & 53.4 & 57.7 \\
\hline 
\hline 
$D_{el,te}$ & MWh\slash a & 29.8 & 29.8 & 29.8 & 29.8 & 29.8 & 36.8 & 36.9 \\
$D_{el,hp}$ & MWh\slash a & - & - & - & 17.2 & 18.8 & 17.2 & 18.6 \\
$D_{el,tot}$ & MWh\slash a & 29.8 & 29.8 & 29.8 & 47.0 & 48.6 & 54.1 & 55.4 \\
\hline 
$cap_{PV}$ & kW & - & - & 5.8 & - & 8.2 & 10.0 & 12.8 \\
$cap_{CHP}$ & kWel & - & 50.0 & 50.0 & 32.8 & 30.8 & 32.7 & 31.1 \\
$cap_{HP}$ & kWth & - & - & - & 11.0 & 11.7 & 11.1 & 11.7 \\
$cap_{WWT}$ & kWh & - & 42.5 & 40.5 & 46.3 & 44.4 & 50.0 & 41.7 \\
$cap_{boil}$ & kWh & 60.3 & - & - & - & - & - & - \\
\hline 
$P_{chp,max}$ & kW & - & 40.1 & 40.1 & 32.8 & 30.8 & 32.7 & 31.1 \\
$h_{chp,full}$ & hours & - & 30188 & 30191 & 30000 & 30000 & 30000 & 30000 \\
\hline 
$SCR_{el}$ & \% & - & 32.7 & 33.3 & 84.8 & 85.0 & 85.9 & 91.0 \\
$DSS_{el}$ & \% & - & 82.7 & 90.1 & 88.9 & 93.4 & 92.3 & 95.4 \\
$DA_{el}$ & \% & - & 253.2 & 270.7 & 104.9 & 109.9 & 107.4 & 104.9 \\
$GII$ & \% & - & 19.5 & 19.6 & 10.6 & 11.4 & 10.5 & 10.0 \\
$GII_{norm}$ & \% & - & 125.5 & 126.6 & 57.6 & 62.2 & 77.7 & 90.5 \\
\hline 
$CO_{2,ref}$ & t & 659.3 & 659.3 & 659.3 & 659.3 & 659.3 & 688.9 & 688.9 \\
$CO_{2,opt}$ & t & 659.3 & 888.4 & 879.3 & 587.5 & 543.5 & 581.0 & 545.8 \\
$\Delta CO_{2}$ & t & - & -229.1 & -219.9 & 71.8 & 115.8 & 107.8 & 143.1 \\
\hline 
$CO_{2,export}$ & t & - & 318.3 & 333.5 & 47.0 & 45.6 & 44.2 & 32.8 \\
$\Delta CO_{2,export}$ & t & - & 105.3 & 107.9 & 15.5 & 12.0 & 9.9 & 10.8 \\
\hline 
$CF_{subs}$ & \kEUR{} & - & 136.8 & 142.5 & 61.7 & 61.9 & 66.0 & 61.7 \\
$cac_{subs}$ & \eurotCO{} & - & - & - & 859.3 & 534.1 & 611.7 & 431.4 \\
\hline 
\end{tabular}
\caption{Results for component wise analysis sorted by $\Delta NPV$, 
                cascading mode with \textbf{triple} the fix price for CHP representing a system with three CHP-units} 
\label{tbl_results_comp_3,0}
\end{table}

\section{PV remuneration}
\label{sec_app_PVremuneration}

To determine the amount of remuneration for PV electricity, the REL 2021 introduces three pricing levels that result in the following subsidies for the 1st of January 2021:
\begin{itemize}
    \item Up to a PV capacity of $\mathit{cap_{pv}}<=10\,\mathit{kW_p}$ the feed-in tariff is \ctkWh{8.56} and the SCP is \ctkWh{3.79}
    \item For a PV capacity of $10\,\mathit{kW_p} < \mathit{cap_{pv}} <= 40\,\mathit{kW_p}$ the feed-in tariff is \ctkWh{8.33} and the SCP is \ctkWh{3.52}
    \item For a PV capacity of $40\,\mathit{kW_p} < \mathit{cap_{pv}} <= 750\,\mathit{kW_p}$ the feed-in tariff is \ctkWh{6.62} and the SCP is \ctkWh{2.37}
\end{itemize}

The prices for the different levels are taken into account proportionally. As an example, Equation \ref{eq_app_PVremuneration} illustrates the resulting feed-in tariff for a PV system of \kWp{50}.

\begin{equation}
    \label{eq_app_PVremuneration}
    \begin{split}
    \mathit{c_{pv,feedin}} &= \Bigg[ 
                            \frac{10}{50} \cdot 8.56 + \frac{40-10}{50} \cdot 8.33 +
                            \frac{50-40}{50} \cdot 6.62 
                            \Bigg] \; \mathit{ct} \slash \mathit{kWh} \\
                            &= 8.03 \; \mathit{ct} \slash \mathit{kWh}
    \end{split}
\end{equation}

For this study to depict that proportional pricing system, we divided the range of possible PV capacities between \kWp{0} and \kWp{100} into 19 remuneration schemes. This is further mentioned in Section \ref{sec.method_OptModel} in Equation \ref{eq.bin_PV} and Equation \ref{eq.cap_PV}, and shown in Table \ref{tbl_PV_remuneration}. For calculating the remuneration scheme, we used the mean capacity of the lower and upper limit of the respective remuneration scheme. For example, for remuneration scheme 10, we calculated the price level for a mean PV capacity of \kWp{52.5}. 

\bibliographystyleSI{elsarticle-num-names}
\bibliographySI{SI_references.bib}

\end{document}